\documentstyle[aas2pp4]{article}

\newcommand{\be}{\begin{equation}}
\newcommand{\ee}{\end{equation}}
\def\jc {J. Comp. Phys.}

\def\w {{\bf w}}
\def\gm {\gamma_1}
\def\hw {\overline{w}}
\def\hP {\overline{P}}

\def\ws {{\tilde \w}}
\def\wss {{\tilde \ws}}

\def\bb {{\bf B}}

\def\uu {{\bf u}}
\def\nn {{\bf n}}

\def\I {{\bf I}}
\def\sB {\tilde {B}}

\def\hb {\overline {B}}

\def\dert {\partial_t}
\def\dere {\partial_{\eta}}
\def\derx {\partial_x}
\def\dery {\partial_y}
\def\xm {x_{j-1/2}}
\def\jm {{j-1/2}}
\def\km {{k-1/2}}
\def\xp {x_{j+1/2}}
\def\jp {{j+1/2}}
\def\kp {{k+1/2}}

\def\ym {y_{k-1/2}}
\def\yp {y_{k+1/2}}

\def\R {{\bf R}}

\def\aA {\hat{A}}
\def\aw {\hat{w}}
\def\aB {\hat{B}}

\def\tuu {\tilde{\uu}}

\def\Ls {\tilde{L}}
\def\Qs {\tilde{Q}}

\def\G {\Omega}
\def\A {{\bf A}}
\def\D {{\bf D}}
\def\ff {{\bf f}}
\def\af {\hat f}
\def\ag {\hat g}
\def\aff {\hat\ff}

\def\gg {{\bf g}}

\def\agg {\hat\gg}
\def\v {{\bf v}}
\def\q {{\bf q}}
\def\dd {{\cal D}^{(2)}}
\def\de {{\cal D}^{(1)}}
\def\divb {\nabla\cdot\bb}
\def\rp {\rho^{\prime}}
\def\prp {p^{\prime}}
\def\qxp {q_{\xi}^{\prime}}
\def\qyp {q_{\eta}^{\prime}}
\def\Byp {B_{\eta}^{\prime}}

\begin{document}

\title{High Order Upwind Schemes for Multidimensional Magnetohydrodynamics}

\author{P. Londrillo}
\affil{Osservatorio Astronomico, Bologna, Italy}
\authoraddr{Via Ranzani 1, 40127 Bologna, Italy \\
e-mail: londrillo@bo.astro.it}
\and
\author{L. Del Zanna}
\affil{Dipartimento di Astronomia e Scienza dello Spazio, Firenze, Italy}
\authoraddr{Largo E. Fermi 5, 50125, Firenze, Italy \\
e-mail: ldz@arcetri.astro.it}

\begin{abstract}
A general method for constructing high order upwind schemes 
for multidimensional magnetohydrodynamics (MHD), having as a main
built-in condition the divergence-free constraint
$\divb=0$ for the magnetic field vector $\bb$, is proposed.
The suggested procedure is based  on {\em consistency}
arguments, by taking into account the specific operator structure
of MHD equations with respect to the reference Euler equations of
gas-dynamics. This approach leads in a natural way to a staggered
representation of the $\bb$ field numerical data where the divergence-free
condition in the cell-averaged form, corresponding to second order
accurate numerical derivatives, is exactly fulfilled. To extend this
property to higher order schemes, we then give 
general prescriptions to satisfy a $(r+1)^{th}$ order accurate
$\divb=0$ relation for any numerical $\bb$ field having a $r^{th}$
order interpolation accuracy.
Consistency arguments lead also to a proper formulation of the
upwind procedures needed to integrate the induction equations, assuring the
exact conservation in time of the divergence-free condition and
the related continuity properties for the $\bb$ vector components.
As an application, a third order code  
to simulate multidimensional MHD flows of astrophysical interest
is developed using ENO-based reconstruction algorithms. Several
test problems to illustrate and 
validate the proposed approach are finally presented.
\end{abstract}

\section{Introduction}

Many astrophysical plasmas, such as stellar (or galactic)
atmospheres and winds, accretion disks and jets,
can be described by the set of compressible magnetohydrodynamic (MHD)
equations with dissipative terms neglected, 
since kinetic effects of astrophysical
plasmas are quite small on dominant macroscopic scales.
In these physical regimes, dynamical effects give rise 
to complex time dependent flows where localized 
sharp modes like shocks and current sheets 
couple with distributed nonlinear waves.
It is therefore a main challenge to computational astrophysics 
to take properly into account both dynamical components. 

Centered finite differences or spectral
schemes are well suited for smooth fields and can support 
discontinuities only by introducing {\em enough}
viscous/resistive dissipation. 
In this way field discontinuities
are represented with a poor resolution and
artificial heating takes place. On the other hand,
upwind schemes achieve shock-capturing and
localized high resolution in a natural way.
When discontinuous solutions are of main interest,
second order (in time and space) schemes are
usually adopted, since they reconcile resolution with efficiency 
and stability needs. But in the general case, when coherent  
sharp field structures are embedded
in a turbulent background, second order accuracy is 
no longer the optimal one, since the (implicit) numerical viscosity
is still to high to resolve properly small scales motions. 
There are then compelling computational and physical reasons to develop
higher order upwind schemes for MHD flows.

In recent years progress has been made in extending
Godunov-type schemes developed for the Euler system of gas-dynamics to MHD,
with main emphasis on the wave characteristic structure. 
In Brio \& Wu (1988),  and Roe \& Balsara (1996) the problem
of non strict hyperbolicity of the MHD system has been addressed
by introducing proper regularity factors to renormalize the
eigenvectors and assure their linear independence.
The related problem of constructing the Roe linearized matrix
for MHD case, has also been solved (Cargo \& Gallice 1997, Balsara 1998a).
Based on these achievements, second order upwind
codes using either Godunov's or Roe's method have then
been constructed and tested, mainly for one-dimensional MHD problems
(e.g. Ryu \& Jones 1995, Zachary et al.~1994, 
Dai \& Woodward 1994, Balsara 1998b).

Specific new problems and
limitations have to be considered in going to
higher order and multidimensional MHD case.
Upwind schemes are usually constructed
by first projecting fluid variables at each grid
point on the space of characteristic variables.
The decomposition
procedure allows to achieve a better resolution since
interacting discontinuities of fluid variables become uncoupled
in the space of characteristic variables. This technique is usually
adopted also in existing MHD codes, but there is no clear evidence
that it can work even for higher order schemes (Barmin et al.~1996).
Moreover, the computational cost to project
field variables onto the MHD seven-component characteristic space
may become prohibitive when moving to higher order and 
higher dimensional schemes. Therefore, as already experienced in
the context of numerical gas-dynamics, a search for high 
order shock-capturing schemes 
where no characteristic decomposition is
needed and where attention is shifted more to
a {\em vanishing viscosity} entropy satisfying 
model equations, rather than on approximate Riemann solvers,
appears to be more promising.

A second important issue in multidimensional MHD schemes
comes from the need to satisfy the divergence-free 
condition of the magnetic field vector. This property
is a crucial one for two main reasons (Balsara \& Spicer 1999):
\begin{itemize}
\item the conservation form of MHD equations for energy and momenta
is based on the implicit $\divb=0$ condition, and
\item all the topological aspects of magnetic field lines
which are relevant to critical MHD phenomena, like reconnection,
heavily rely on this condition.  
\end{itemize}
On the other hand, this specific property has no easy representation
in a numerical framework, like 1-D Godunov-type schemes,
designed to handle compressive modes and shocks.
This longstanding problem has been addressed by many authors
and several recipes have been proposed and experimented so far.
Depending on the adopted methodology, these works can be
broadly classified into three main categories:
\begin{enumerate}
\item Many MHD codes are constructed by simply extending 
to higher dimensions 1-D Riemann solvers using a directional
splitting technique, as for the Euler system
(Zachary et al.~1994, Ryu et al.~1995,
Balsara 1998b, for second order Godunov-type schemes;
Jiang \& Wu 1999, for $5^{th}$ order WENO scheme).
In this approach the $\divb=0$ condition
breaks down, of course, and some correction step has then to be
applied. Following Brackbill \& Barnes 1980, a
cleaning procedure is usually carried out by solving
a Poisson equation, which is 
equivalent to add a new (elliptic) equation to
the original hyperbolic MHD system. As an empirical recipe, this
method is by no means optimal and may lead to inconsistencies.
In particular, the numerical derivatives appearing in 
Poisson equation have no clear relation with the upwind
derivatives of the base MHD system and the boundary conditions 
become indeterminate for nontrivial boundary-value problems.

\item In the Powell (1994) approach it is first pointed out 
the formal difficulty of applying 1-D Riemann solvers to the multidimensional
case, since the 1-D
MHD mode eigenspace, having seven components, is not of full
rank for the 2-D case, where an eight-component state vector
is involved. Therefore, variations of magnetic field components
appearing in the $\nabla\cdot {\bf B}$ operator cannot be represented 
by Riemann solvers based on the 1-D eigenspace.
This undoubtedly correct premise led the author to 
propose a modification of the MHD equations by adding 
a ``$\nabla\cdot {\bf B}$ mode'', propagating with the local flow speed.
In this way the hyperbolic character of the MHD system is surely retained,
at the price of suppressing the divergence-free property. This approach 
appears to be highly questionable, of course, since important
physical properties of the MHD equations, and especially
magnetic field topologies, are clearly lost.

\item In the present work we have taken as a main
starting point all those references attempting to design upwind
schemes where a numerical divergence-free condition works as a build-in
property (Evans \& Hawley 1988, DeVore 1991,
Stone \& Norman 1992, Dai \& Woodward 1998, Ryu et al.~1998,
Balsara \& Spicer 1999, among others).
In all these works, the introduction of the magnetic
vector potential or the equivalent conservative formulation of Stoke's
theorem lead to represent magnetic field components at 
staggered collocation points, and a numerical
$\divb=0$ relation follows as an algebraic identity.
Moreover, when induction equations are properly formulated in terms
of the staggered fields, conservation in time of the divergence-free
property is also assured.

In the cited works, however, some main questions are still left
open. These are essentially related to a 
persisting {\em duality} between staggered
magnetic components evolving in the induction equations and the
same components, now collocated at node points (or cell centers) as 
other fluid variables, entering the Riemann solver procedures.
Several recipes based on interpolation 
have been suggested to relate cell centered and staggered fields.
However, as widely discussed in the
Dai \& Woodward (1998) paper, the cell centered field components
do not preserve, in general, the original divergence-free property, 
unphysical magnetic monopoles still arise and their sizes seem to
depend on the adopted interpolation schemes.

A related question concerns how upwind fluxes
in the induction equations have to be formulated, since
1-D Riemann solvers for 
density, momentum and energy equations have no straightforward
extension to them, when staggering is adopted. 
Again, many different empirical solutions
have been proposed, which hardly can be compared and evaluated
as long as only qualitative numerical tests are at disposal.

\end{enumerate}

In the present paper we propose some {\em general} answers
to these questions, by showing that 
{\em consistency} arguments are sufficient to envisage the
main rules to adapt upwind schemes designed for Euler equations
to the MHD case. Consistency requires that
the specific operator structure of the MHD system and 
the related magnetic field properties have to be preserved
by discretized equations and upwind procedures. 
In this way, different schemes and
their high order extensions can be designed,
all assuring a numerical divergence-free condition as well as
the related uniqueness and regularity of magnetic field lines.
On the same ground, existing MHD codes and published
numerical results can be evaluated on a more appropriate framework.

The plan of the paper is as follows. In Sect.~2 the general formulation
to discretize Euler and MHD equations in conservation form is 
reviewed, with emphasis on the differences in space operator
structures and on the related numerical representation.
In Sect.~3, the kinematical properties
of discontinuous, divergence-free magnetic field are 
first analyzed  to represent numerical data and then 
used  to construct appropriate 
upwind flux formulas (or approximate Riemann solvers) for the 
MHD equations. The proposed formulation is then also
compared to recently published schemes.
In Sect.~4 a code for 2-D systems,
based on third order ENO-type reconstruction procedures and on
the simple Lax-Friedrichs flux upwinding, is presented.
Sect.~5 is devoted to numerical test problems to
add confidence and validation of the proposed approach and
conclusive remarks are briefly given in Sect.6.

\section{Euler versus MHD systems}

To underline differences between the Euler
and MHD systems relevant to numerical discretization,
we first briefly review some of the main points characterizing
upwind schemes for the Euler equations. As a general 
framework, we consider here the flux vector splitting (FVS) formalism
(van Leer 1982, Chen \& Lefloch 1995) and the high order reconstruction
techniques based on polynomials (Shu 1997). For ease of
presentation we treat only the 2-D case in cartesian geometry, 
being both the 3-D case and curvilinear geometries just straightforward
extensions. 

\subsection{Upwind schemes for Euler equations}

The equations of gas-dynamics in two spatial dimensions
constitute a system of $m=5$ conservation laws:
\be 
\dert\uu+\derx [\ff(\uu)]+\dery [\gg(\uu)]=0,
\label{euler}
\ee
where $\uu=[\rho,\q,e]^T$
is the state vector of conservative variables, and
$$\ff=[q_x,v_xq_x+p,v_xq_y,v_xq_z,v_x(e+p)]^T,$$
$$\gg=[q_y,v_yq_x,v_yq_y+p,v_yq_z,v_y(e+p)]^T,$$
are the corresponding flux vector functions. 
Here $\rho$ is the mass density, $\q=\rho\v$ the momentum
associated with the flow velocity $\v$, $e$ the total energy per unit volume
and $p=(\gamma-1)[e-\q\cdot\v/2]$ the gas pressure for a $\gamma-$law 
equation of state.

A basic property of the Euler system is that each Jacobian matrix,
$\A_x(\uu)=\partial_{\uu}\ff(\uu)$ and $\A_y(\uu)=\partial_{\uu}\gg(\uu)$,
has a set of $m$ real eigenvalues $\{\lambda^{s}(\uu)\}~(s=1,2,\dots,m)$, 
and a corresponding complete set of right $\{\R_s(\uu)\}$
and left $\{\R^{-1}_s(\uu)\}$ eigenvectors, at every point $\uu$
(hyperbolicity properties). Physically relevant solutions to
system~(\ref{euler}) are selected by imposing the admissibility condition
\be
\dert[\rho F(s)]+\nabla\cdot[\rho F(s)\v]\le 0,
\label{entropy}
\ee
where $F(s)$ is any smooth function of the
specific entropy $s(p,\rho)$ (Harten et al.~1998).

Numerical schemes for system~(\ref{euler}) use the following 
{\em consistency conditions} as general guidelines (Tadmor 1988):
\begin{itemize}
\item the conservation form, assuring that numerical solutions
capture correctly weak solutions;
\item the entropy inequality, to be preserved by the discretized
entropy functions. Upwind schemes are then designed to
have (implicit) numerical viscosity
compatible with relation~(\ref{entropy}).
\end{itemize}
In the semi-discrete formulation, appropriate for higher $r>2$
order schemes, space operators are approximated 
(for a fixed time $t$) on a $N_x\times N_y$
dimensional grid with node points $N_{j,k}\equiv (x_j,y_k)$, where
$x_j=h_xj\,(j=0,1,\dots,N_x-1)$ and $y_k=h_yk\,(k=0,1,\dots,N_y-1)$;
here $h_x$ and $h_y$ are the constant grid sizes along each direction.
The point values formulation based on $\{\uu_{j,k}\}$
data leads then to the conservative scheme
$${d\uu_{j,k}(t)\over dt}=-{1\over h_x}[\aff_{j+1/2,k}-
\aff_{j-1/2,k}]$$
\be
-{1\over h_y}[\agg_{j,k+1/2}-\agg_{j,k-1/2}],
\label{euler2}
\ee
where $\aff_{\jp,k}$ and $\agg_{j,\kp}$ denote the numerical
flux-vector functions needed 
to approximate the corresponding flux derivatives to a given order $r$. 
A numerical approximation is then characterized essentially 
by the way $\aff_{\jp,k}$ and $\agg_{j,\kp}$ are evaluated 
for a given set of $\{\uu_{j,k}(t)\}$ data. Time integration can then be 
performed by appropriate Runge-Kutta or equivalent
stable discretization schemes (Shu \& Osher 1988).

Modern higher order shock-capturing schemes
generalize first order Godunov scheme 
by following two main steps (Harten et al.~1987):
\begin{enumerate}
\item a reconstruction phase to recover variable values
at grid points where flux derivatives have to be computed;
\item an upwind phase, where the Godunov method for a scalar
conservation law in one dimension is extended to the $m>1$
components system in higher dimensions.
\end{enumerate}
As far as item~1 is concerned, we remind here some basic points relevant
to the following sections and to the actual code structure, to be presented
in Sect.~4.

Any one-dimensional piecewise smooth function $w(x)$,
defined by cell averaged data 
$$\hw_j={1\over h_x}\int_{\xm}^{\xp}w(x)dx,$$
can be approximated by uniform $(r-1)^{th}$ order
piecewise polynomials $P_j(x;\hw)$, which must have the conservative 
property $\hP_j=\hw_j$. Likewise, a $w(x)$ function defined by
grid-point data $\{w_j=w(x_j)\}$ can be approximated by 
interpolation polynomials $P_j(x;w)$ defined as $P_j(x_j;w)=w_j$.
Here we denote with $R[x;\hw]$ (or $R[x;w]$) the corresponding
polynomials set $\{P_j\}$ to reconstruct $w(x)$ at any $x$ point.

At points of discontinuity, $R[x;\cdot]$ has to
satisfy definite non-oscillatory constraints, for accuracy and
stability purposes. Standard references are provided by
linear polynomials based on {\em minmod} limiters to preserve monotonicity
of data (MUSCL scheme: van Leer 1979; TVD scheme: Harten 1983)
or by higher $r>2$ order polynomials based on ENO procedures
(Harten et al.~1987, Shu \& Osher 1989) having
weaker (essentially non-oscillatory) monotonicity properties.
Piecewise polynomials approximate the $w(x)$ function 
at any cell boundary point $\xp$  by a  
two-point left-right $(w^{(L)},w^{(R)})$ value, where
$$[w^{(L)}]_\jp=P_j(\xp)=w(\xp^-)+O(h_x^{r}),$$
$$[w^{(R)}]_\jp=P_{j+1}(\xp)=w(\xp^+)+O(h_x^{r}),$$
for smooth functions.
For $w(x)$ having a discontinuous $k^{th}$ 
derivative in the $x_j < x< x_{j+1}$ range, with $k< r$,
the  accuracy order becomes $O(h_x^{k+1})$.

The reconstruction procedures can be extended to the 2-D functions
in Eq.~(\ref{euler}) by assuming that the scalar components
$u(x,y)$ (and hence $f[\uu(x,y)]$ and $g[\uu(x,y)]$)
are piecewise smooth along each coordinate. 
In this way, calling $C_{j,k}$ the 2-D cell centered at the
node point $N_{j,k}$, the functions $u(x,y)$
can be reconstructed at the cell edges $(\xp,y_k)$
and $(x_j,\yp)$ by using respectively the 1-D operators
$R[x;u_k]$ and $R[y;u_j]$. The same functions may also be
reconstructed at a cell corner $P_{j,k}\equiv (\xp,\yp)$ 
by using the 2-D compound operators $R[\xp;R[\yp;u]]$.
Actually, space discretization in Eq.~(\ref{euler})
involves only 1-D reconstructions, one for each direction.
In fact, $\af_{\jp,k}$ is defined at the point $x=\xp$, for
fixed $y=y_k$, where the argument variables have
the two-state $x-$wise reconstructed values $[u^{(L,)},u^{(R,)}]_k$;
likewise, $\ag_{j,\kp}$ is defined at the point $y=\yp$, 
for fixed $x=x_j$, where the argument variables have the
two-state $y-$wise reconstructed values $[u^{(,L)},u^{(,R)}]_j$.

Let us now turn our attention to item~2.
In the FVS formalism, to represent a flux variation, say 
in the $x$ coordinate, the vector $\ff(\uu)$ is decomposed as
$$\ff(\uu)={1\over 2}[\ff^{(+)}+\ff^{(-)}],\quad
\ff^{(\pm)}=\ff(\uu)\pm \D_x(\tuu)\cdot\uu,$$
for states $\uu$ around a given reference (constant) state
$\tuu$. The flux vectors $\ff^{(\pm)}$ have Jacobian matrices 
$\A_x(\tuu)\pm\D_x(\tuu)$ with only positive/negative eigenvalues.
The matrix $\D_x$ is then required to be real (symmetrizable)
and positive, $\D_x(\tuu) \ge |\A_x(\tuu)|$, where
$$|\A_x(\tuu)|=\sum_s[\R_s|\lambda_s|\R^{-1}_s]_{\tuu}\,.$$
For given $\uu_1=\uu^{(L,)}$,\,$\uu_2=\uu^{(R,)}$ reconstructed values
at the point $\xp$ and for fixed $k$ index, flux splitting allows to
define the numerical flux
$$\ff(\uu_1,\uu_2)_{\jp,k}=$$
\be
{1\over 2}[\ff(\uu_1)+\ff(\uu_2)-
\D_x(\tuu)\cdot(\uu_2-\uu_1)]_{\jp,k},
\label{f}
\ee
which has {\em upwind properties}, namely 
it is a two-point vector function
non-increasing in the first argument and
non-decreasing in the second argument.
The reference state $\tuu$ is given by the Roe average
or by the simpler $\tuu=(\uu_1+\uu_2)/2$ arithmetic average, in a way
to assure consistency $\ff(\uu,\uu)=\ff(\uu)$ and
continuous dependence on data.

Standard references for the scheme in Eq.~(\ref{f}) are given by 
either the approximate Riemann solvers of Roe-type
or of Godunov-type , where 
the corresponding matrix $\D_x$ has the formal property
$$\D_x(\tuu)=|\A_x(\tuu)|+O(|\uu^{(L,)}-\uu^{(R,)}|)$$
entailing a minimum of numerical viscosity compatible with the entropy law.
On the other hand, a maximum of numerical
viscosity is achieved by the (global) Lax-Friedrichs (LF) flux, 
where the $\D_x$ matrix reduces to the simple diagonal form
$$\D_x(\tuu)=\alpha\I,\quad
\alpha=max_{\tuu}[max_s|\lambda_s(\tuu)|],$$
and in which Riemann characteristic informations are averaged out.
Note that the flux formula Eq.~(\ref{f}) can be interpreted either as an
approximate Riemann solver based on local linearization, or as a 
discrete approximation of the associated viscosity model equation. In fact,
a first order approximation of the flux splitting is the discretized
representation of the viscous flux
\be \ff_v(\uu)=\ff(\uu)-h_x\D_x\cdot\derx\uu,
\label{fv}
\ee
which provides a link between the entropy condition and the numerical
viscosity associated to the dissipation matrix $h_x\D_x$.

In 2-D problems,
the semi-discrete formulation of Eq.~(\ref{euler2}) 
and the independence of the Jacobian
matrices $(\A_x,\A_y)$, allow one to represent the numerical fluxes
$\aff_{\jp,k}$ (for fixed $y_k$) and $\agg_{j,\kp}$
(for fixed $x_j$) through independent upwind procedures,
constructed with the matrices $\D_x$ and $\D_y$, respectively. 
In this way, the flux formula for $\gg$ is given by
$$\gg(\uu_1,\uu_2)_{j,\kp}=$$
\be
{1\over 2}[\gg(\uu_1)+\gg(\uu_2)-
\D_y(\tuu)\cdot(\uu_2-\uu_1)]_{j,\kp},
\label{g}
\ee
where $\uu_1=\uu^{(,L)}\,$ $\uu_2=\uu^{(,R)},$
are the reconstructed values at $y=\yp$, for fixed $x=x_j$.
The corresponding viscosity form of the numerical flux $\gg$
is given by
\be \gg_v(\uu)=\gg(\uu)-h_y\D_y\cdot\dery\uu.
\label{gv}
\ee
The construction of upwind fluxes in Eqs.~(\ref{f}) and (\ref{g}),
based on the $\A_x$ matrix eigenspace at the point $(\xp,y_k)$
and on the $\A_y$ matrix eigenspace at the point $(x_j,\yp)$,
is usually referred to as a {\em directional splitting} setting. 
This procedure is consistent with the divergence
form of space operators in Eq.~(\ref{euler}) and implies that
the space derivatives are obtained by summing the two flux differences
computed both at the same time $t$.
In the Strang-type formalism, which is widely adopted in the numerical
astrophysics community, a splitting procedure is also applied to the
time evolution operators, by constructing the updated solution
of a 2-D problem as a sequel of independent 1-D problems, one
for each direction, in turn. In the Euler system directional
splitting or time splitting procedures give (formally)
equivalent results, at least for second order schemes.
In the MHD case, however, this
formal equivalence is definitely lost, as it will be discussed
in the following.

\subsection{The two-dimensional MHD system}

The set of MHD conservation laws cannot be considered as a simple
extension of the Euler system, with just a higher number
of state variables. Actually, while the conservation form,
the entropy law and the general hyperbolic properties are maintained,
some specific differences related to the structure of the space operators
have to be considered.

In fact, the MHD system can be viewed as composed by two coupled
subsystems, the first one containing space operators in
the divergence form as in Eq.~(\ref{euler}) evolving density energy
and momenta, and the second
one, specific to the magnetic field evolution, containing space operators
in the {\em curl} form. In both subsystems the $\divb=0$ property
of the vector magnetic field enters in a substantial way and has then
to be considered as a new consistency condition for numerical discretization.  

By specializing again to 2-D systems, the MHD equations are given by
\be
\dert \uu+\derx [\ff(\w)]+\dery [\gg(\w)]=0,
\label{MHDu}
\ee
for the six-component state vector $\uu=[\rho,\q,e,B_z]^T$,
coupled with the induction equations 
\be
-\dert B_x+\dery \Omega(\w)=0,\quad\dert B_y+\derx \Omega(\w)=0,
\label{MHDb}
\ee
for the (poloidal) vector field $\bb=[B_x,B_y]^T$.
We denote the overall eight-component state vector as $\w=[\uu,\bb]^T$.
The flux vectors in Eq.~(\ref{MHDu}) are given by
$$\ff(\w)=[q_x,F_{x,x},F_{x,y},F_{x,z},E_x,G_{x,z}]^T,$$
$$\gg(\w)=[q_y,F_{y,x},F_{y,y},F_{y,z},E_y,G_{y,z}]^T,$$
where, for indexes $i,j=x,y,z$:
$$F_{i,j}=v_iq_j-B_iB_j+\Pi\delta_{i,j},$$
$$E_i=v_i(e+\Pi)-B_i(v_jB_j),$$
$$G_{i,j}=v_iB_j-v_jB_i,$$
in which the relations $F_{i,j}\equiv F_{j,i}$ and $G_{i,j}\equiv-G_{j,i}$
clearly hold.
Here $\Pi=p+(B_iB_i)/2$ and $p=(\gamma-1)[e-(q_iv_i)/2-(B_iB_i)/2]$ 
are respectively the total and the gas pressures.
The common flux function of Eqs.~(\ref{MHDb}) is defined
as $\Omega=G_{x,y}\equiv-G_{y,x}=v_xB_y-v_yB_x$.

For given $\bb$ field, the subsystem~(\ref{MHDu}) as an {\em Euler form}, with
independent Jacobian matrices of full $m=6$ rank, so that the upwinding
procedures based on directional splitting of the previous section
can be extended. Differences arise, however, for the induction equations. 
In fact, subsystem~(\ref{MHDb}) is generated by a {\em unique} 
flux function and 
has then only a one-dimensional eigenspace.
This formal
property is clearly related to the $\divb=0$ condition, as can be
better evidenced by introducing a vector potential representation
of the (poloidal) $\bb$ components (here $A\equiv A_z$)
\be
B_x=\dery A,\quad B_y=-\derx A,
\label{b}
\ee
For given smooth $(B_x,B_y)$ fields, $A(x,y)$ always exists as a 
one-valued differentiable function. For discontinuous
fields, Eq.~(\ref{b}) still holds in weak form, implying that 
$A(x,y)$ is at least (Lipschitz) continuous along each coordinate.
On the other hand, for given $A(x,y,t)$, the system given by 
Eqs.~(\ref{MHDb}) is fully equivalent to the one-component evolution equation
\be
\dert A-\Omega(\w)=0,
\label{Az}
\ee
coupled with Eqs.~(\ref{b}).

In a formal setting, for given state 
variables $\uu$, Eq.~(\ref{Az}) is a Hamilton-Jacobi
equation (Jin \& Xin 1998), and Eqs.~(\ref{MHDb}) represent
the associated hyperbolic system.
An important property of Eqs.~(\ref{Az}) is that $A(x,y,t)$
is continuous at all times and only discontinuities
in its first derivatives may develop. Therefore, field lines defined
by the isocontours $A(x,y)=\mbox{const}$ are allowed to have {\em corners},
but not jumps.
 
The overall MHD system can then be viewed as a coupled system
of a Hamilton-Jacobi equation and of a set of conservation laws
in the Euler form. One consequence is that  the Jacobian matrix
$\A_x$ corresponding to the $[\ff,\Omega]^T$ vector flux
is only of $m=7$ rank and can represent characteristic modes 
of variables $\w_x=[\uu(x,),B_y(x,)]^T$, while the independent matrix
$\A_y$, corresponding to the $[\gg,-\Omega]^T$ vector flux,
can represent variables $\w_y=[\uu(,y),B_x(,y)]^T$. It is evident
that the missing degrees of freedom
$[B_x(x,),B_y(,y)]$ are not evolutionary and
cannot have a characteristic-based representation.

A numerical schemes preserving these general 
properties must then be be characterized by the following points:
\begin{enumerate}
\item A numerical $\divb=0$ condition and its conservation in time
imply that the induction equations~(\ref{MHDb})
have to be discretized using a {\em unique} flux function $\Omega(\w)$
located at common points. This entails necessarily a staggered 
collocation of the magnetic field scalar components.
\item A divergence-free magnetic field is fully equivalent
to its representation via a numerical vector potential and
likewise the evolution equations~(\ref{MHDb}), discretized as in
item~1, can always be integrated via the scalar Eq.~(\ref{Az}).
\item Relevant to the reconstruction and upwind steps
is that the magnetic field components are at least continuous
along the respective longitudinal coordinates, while 
discontinuities, to be related to the MHD characteristic modes,
can occur only along the respective orthogonal coordinates (see below).
\end{enumerate}

\section{$\divb=0$ preserving upwind schemes for MHD equations}

In this section we concentrate on the correct collocation
and reconstruction step 
for the numerical magnetic field data, and then on the upwind
flux formulation for the induction equations, in order to preserve
the peculiar features of the MHD system as outlined just above.

\subsection{The reconstruction step}

While for given data $\{u_{i,j}\}$ of $\uu$ variables in Eq.~(\ref{MHDu})
the reconstruction procedures follow 
the same lines as in the Euler system~(\ref{euler2}), 
for the field $\bb$ it is necessary to take into account
the $\divb=0$ condition as a new kinematical constraint.
For general piecewise smooth fields this condition is 
expressed in integral form by
\be
\int_{\partial C} B_n d\,l =0 \label{int}
\ee
for any cell $C$, where
$B_n=\bb\cdot\nn$ and $\nn$ is the unit vector normal to 
the boundary line $\partial C$. 
By first choosing a cartesian cell with 
sides $[2\epsilon,h_y]$, the following continuity condition
for the $y-$averaged $\hb_x(x)$ component at any point $x$ comes out:
$$\hb_x(x+\epsilon)-\hb_x(x-\epsilon)=O(\epsilon).$$
The same argument leads to the continuity of the $x-$averaged
$\hb_y(y)$ field at any point $y$.

If Eq.~(\ref{int}) is then integrated 
over a computational cell $C_{j,k}$, one has 
$$h_y[\hb_x(\xp)-\hb_x(\xm)]_k+$$
\be
h_x[\hb_y(\yp)-\hb_y(\ym)]_j=0,
\label{div}
\ee
where $[\hb_x]_k$ is the $y$ average on the vertical cell side centered on
$y_k$ and $[\hb_y]_j$ the corresponding $x$ average over the horizontal
cell side centered on $x_j$. 

Continuity conditions and Eq.~(\ref{div})
are thus the main ingredients to construct at any point
a divergence-free numerical field.
In particular, the continuity condition
$\hb_x(x^{+})=\hb_x(x^{-})$ allows one to locate
the $[\hb_x(x)]_k$ field at cell boundary points $\{\xp\}$, where 
all the other variables $[\w_x(\xp)]_k$ are represented by
reconstructed two-state (left-right) values. This property can be expressed
in a formal way by setting 
$[\hb_x^{(L,)}]_{\jp,k}=[\hb_x^{(R,)}]_{\jp,k}$. 
Correspondingly, 
$\hb_y(y)$ can be located at cell boundary points 
$\{\yp\}$ and point values can be interpreted as  
$[\hb_y^{(,L)}]_{\kp}=[\hb_y^{(,R)}]_{\kp}$.

The reconstruction step for the $(B_x,B_y)$ fields
along the respective transverse coordinates leads to the point values
$$ [B_x(y)]_{\jp}=R[y;\hb_x],\quad[B_y(x)]_{\kp}=R[x;\hb_y],$$ 
which have relevance for upwind computations.
In fact, at the $\yp$ point, $B_x(\yp) = [ B_x^{(,R)}, B_x^{(,L)} ]$
is a two-state variable, and can have only a transverse discontinuity line
with a $\delta_yB_x=[B_x^{(,R)}-B_x^{(,L)}]$ jump.
Similar arguments apply to the reconstructed values
$B_y(\xp)=[B_y^{(R,)},B_y^{(L,)}]$ at the $x=\xp$
boundary point, allowing only for a transverse discontinuity
$\delta_xB_y=[B_y^{(R,)}-B_y^{(L,)}]$.

As already noticed by Evans \& Hawley (1988), 
a main property related to condition~(\ref{div})
is that the numerical data $\{\hb_x,\hb_y\}$
allow one to construct a unique (continuous)
numerical potential
$A(\xp, \yp)$, located at the cell corners
$P_{j,k}$. 
It is evident that also the reverse condition holds true,
by first introducing a 
numerical continuous function $A(\xp,\yp)$
and then defining the averaged magnetic field components through Eqs.~(\ref{b}):
$$[\hb_x(\xp)]_k={1\over h_y}[\Delta_y A(\xp)]_k,$$
\be
[\hb_y(\yp)]_j=-{1\over h_x}[\Delta_x A(\yp)]_j, 
\label{bb}
\ee
where $(\Delta_x,\Delta_y)$ denote the usual (undivided)
centered finite differences on the first and 
second coordinate index, respectively.
In this way the divergence-free condition, Eq.~(\ref{div}), 
is identically satisfied by the commutativity of 
the $\Delta_x$ and $\Delta_y$ linear operators, 
whereas the continuity conditions follow from the 
continuity of $A(x,y)$.
These arguments show, in particular,
that the staggered collocation for $\{\hb_x,\hb_y\}$
data is by no means a numerical trick but arises in a consistent
way from Eq.~(\ref{div}) or Eqs.~(\ref{bb}). 

Equation (\ref{div}) gives the cell average of the $\divb=0$
condition, and thus it is an exact law for second order
accurate schemes since $B_x=\hb_x+O(h_y^2)$ at any point $(\xp,y_k)$
and $B_y=\hb_y+O(h_x^2)$ at the corresponding staggered point.
Now, if higher order approximation of point-valued
$B_x$ and $B_y$ fields were recovered using independent reconstruction steps
based on $\hb_x$ and $\hb_y$ data, a numerical $\divb$ of any size
could arise, in
general, since 1-D reconstruction operators do not commute.
To overcome this main difficulty, a different strategy has to be
adopted, by first reconstructing accurate first derivatives based on the
vector potential representation and having a numerical $\divb=0$
relation as a build-in property.
As an illustration, in the following we consider third order
interpolations, but extensions to higher order can be easily
pursued.

We first notice that cell averaged $\hw_j$ and point values
$w_j$ data of a given $w(x)$ function are related by
$\hw_j=w_j +\gm\dd_x(w)_j+O(h^3)$, where $\gm=1/24$ and $\dd_x$
denotes a nonoscillatory numerical second derivative along the $x$
coordinate. To the same accuracy, the inverse relation
$w_j=\hw_j -\gm\dd_x(\hw)_j$
approximates point values using averaged data.
This algorithm, now applied to the values
$w_{\jp}$ at cell interfaces,
$\aw_{\jp}=[w -\gm\dd_x (w)]_{\jp}$,
gives the numerical $\aw(x)$ primitive function
whose two-point difference
$[\Delta_x\aw]_j/h_x$  constitutes a third order approximation of the
$\derx w(x)$ first derivative at $x_j$. 
Let then apply this recostruction step to approximate the 
primitive $\hat {A}$ of the
magnetic potential $A(\xp,\yp)$.
In the 2-D $(x,y)$ plane we have
$$[\hat {A}]_{\jp,\kp}=[A-\gm(\dd_x+\dd_y)A]_{\jp,\kp},$$
and the numerical magnetic field components are
$$[\aB_x]_{\jp,k}={1\over h_y}[\Delta_y\hat {A}]_{\jp,k},$$
\be
[\aB_y]_{j,\kp}=-{1\over h_x}[\Delta_x\hat {A}]_{j,\kp}.
\label{prim}
\ee
By definition, the difference $[\Delta_x\aB_x]/h_x$
gives a third order approximation of the $\derx B_x$ first
derivative at the node $(x_j,y_k)$ point, and 
$[\Delta_y\aB_y]/h_y$ gives the corresponding approximation
of the $\dery B_y$ derivative with the same accuracy and at the
same point. We notice that no left or right derivatives are defined
along the longitudinal coordinates, thus these
numerical approximations are unique. Moreover, one easily
verifies that $\divb=0$, now in the point-valued
form, is exactly fulfilled due to the commutativity of the
$\Delta_x\Delta_y$ operator. The key point here is that to higher
orders only the primitives $[\aB_x(x,),\aB_y(,y)]$ can be
reconstructed directly using a common magnetic potential function
$\aA(x,y)$, but {\em not} the $(B_x,B_y)$ functions themselves
(those entering the fluxes, where divergence-free fields are actually needed).
To achieve this, one needs a further computational step, that is to solve the
(now implicit) relations
$$[B_x-\gm\dd_x B_x]_{\jp,k}=[\aB_x]_{\jp,k},$$
\be
[B_y-\gm \dd_y B_y]_{j,\kp}=[\aB_y]_{j,\kp},
\label{pval}
\ee
where on the left hand sides appear the primitives, and hence the derivatives,
defined in terms of the (unknown) field point values, while on the
right hand sides are the source terms $(\aB_x,\aB_y)$, given by
Eqs.~(\ref{prim}). The numerical fields $(B_x,B_y)$ defined  by these
equations and the corresponding (longitudinal) derivatives are
third order approximations but 
satisfy the divergence-free condition exactly.
In practice one can solve
Eqs.~(\ref{pval}) by some explicit iterative algorithm, 
since each operator $(1-\gm\dd)$ is clearly invertible and a
fourth order accurate, at least, $\divb=O(h_x^4,h_y^4)$ condition 
can then be easily satisfied (see Sect.~4).

This completes the main proof for the reconstruction step, needed to
represent magnetic field  point values in the momentum and
energy equations, where longitudinal
derivatives and hence a $\divb=0$ condition has to be satisfied
to avoid numerical monopoles. Moreover, for given
$[B_x]_{\jp,k}$ and $[B_y]_{j,\kp}$ data it is possible to get interpolated
values at other collocation points, where these field
components act as independent variables and no divergence-free condition
is then required.

\subsection{The upwind procedures}

Let us now consider the first set of MHD equations~(\ref{MHDu})
discretized as in Eq.~(\ref{euler2}):
$${d\uu_{j,k}(t)\over dt}=$$
\be
-{1\over h_x}[\aff_{j+1/2,k}-
\aff_{j-1/2,k}]-{1\over h_y}[\agg_{j,k+1/2}-\agg_{j,k-1/2}],
\label{euler3}
\ee
where now the flux functions $[\ff(\w),\gg(\w)]$
depend on the eight-component vector $\w=[\uu,\bb]$.
The upwind flux based on the $\A_x$ characteristic eigenspace
has the form (consult Eq.~(\ref{f})):
$$[\ff(\w_x,B_x)]_{\jp,k}=
{1\over 2}[\ff(\w_x^{(R,)},B_x)+\ff(\w_x^{(L,)},B_x)$$
\be
-\D^{(1-6)}_x(\ws)\cdot(\w_x^{(R,)}-\w_x^{(L,)})]_{\jp,k}
\label{f2}
\ee
where now upwind properties involve only the variables 
$\w_x=[\uu,B_y]$. 
In the same way,
the upwind flux based on the $\A_y$ characteristic eigenspace
has the form (consult Eq.~(\ref{g})):
$$[\gg(\w_y,B_y)]_{j,\kp}=
{1\over 2}[\gg(\w_y^{(,R)},B_y)+\gg(\w_y^{(,L)},B_y)$$
\be
-\D^{(1-6)}_y(\ws)\cdot(\w_y^{(,R)}-\w_y^{(,L)})]_{j,\kp},
\label{g2}
\ee
where this time the upwind properties involve only
the variables $\w_y=[\uu,B_x]$.

To second order approximation, the numerical flux needed to compute 
space derivatives in 
Eq.~(\ref{euler3}) are given by the flux values 
of Eqs.~(\ref{f2},\ref{g2}), whose arguments
$\w_x$ and $\w_y$ are second order interpolated variables.
In particular $B_x=\hb_x)$ in the $\ff(\w_x,B_x)$ 
flux and $B_y=\hb_y$ in the $\gg(\w_y,B_y)$ flux.

On the other hand, in classical second order schemes for the Euler
equations the numerical flux $\aff$ is 
expressed using cell centered variables as
$$\aff_{\jp,k}={1\over 2}[\ff(\w_{j+1,k})+\ff(\w_{j,k})]$$
$$-{1\over 2}[\D^{(1-6)}_x(\ws)\cdot(\w_x^{(R,)}-\w_x^{(L,)})]_{\jp,k}$$
and in a similar way for
the $\gg$ flux. However, this flux representation cannot be extended to
the MHD case  since cell centered $(B_x,B_y)$ fields are not related
by a divergence-free condition.
The same remark applies to higher $r>2$ order schemes 
where the numerical flux reconstruction is based on cell centered values
(Shu \& Osher 1989).

A second consequence of MHD structure in the
evolution equations~(\ref{euler3}) is that
the $\uu$ state vector has to be integrated in time by {\em summing
flux derivatives evaluated at the same time} $t$, when the implicit
$\divb=0$ condition holds. In those schemes where time integration is
performed by a Strang-type splitting procedure, flux derivatives and hence
terms containing $\derx B_x$ and $\dery B_y$ are necessarily summed at different
time steps and the required $\derx B_x+\dery B_y$ cancellation never occurs.

To summarize, higher order upwind schemes developed for Euler
equations can be extended to the MHD sub-system~(\ref{euler3}), 
with the proviso
\begin{enumerate}
\item flux derivatives have to be computed using flux values
and hence $\bb$ field data {\em directly} collocated at staggered
(i.e.~cell boundary centered) points and not
at the cell centered points and,
\item the same derivatives along the two directions have to be computed
{\em at the same time}, thus avoiding time-splitting techniques.
\end{enumerate}

As already anticipated, the second set of MHD equations, given by the
induction equations~(\ref{MHDb}) for the magnetic poloidal field
components, needs a particular treatment. In the proper, divergence-free
preserving discretized form, these equations are given by
$${d\over dt}[\hb_y(t)]_{j,\kp}=-{1\over h_x}[\Delta_x \Omega(\w_{P})],$$
\be
{d\over dt}[\hb_x(t)]_{\jp,k}={1\over h_y}[\Delta_y \Omega(\w_{P})],
\label{ind}
\ee
where  now $\w=[v_x,v_y,B_x,B_y]^T$
denotes only the variables which are arguments of $\Omega$. In
a fully equivalent form using the vector potential representation, one has 
\be
{d\over dt}[A(t)]_{P}=\Omega(\w_{P}),
\label{dAz}
\ee
to be coupled with Eqs.~(\ref{bb}). 
In both formulations a common flux function
$\Omega(\w)$, located at the cell corner point $P=[\xp,\yp]$, has
to be evaluated.

In order to single out a consistent numerical flux function 
in Eq.~(\ref{dAz}), one has to take into account that
$\Omega(\w_P)$ is now a four-state function
$$\Omega^{(a,b)}=\Omega[\w^{(a,b)}],\quad a,b=R,L,$$
and upwind rules involve necessarily {\em both} $\A_x(\ws)$ and
$\A_y(\ws)$ matrix eigenspaces evaluated at a common reference
state $[\ws]_{P}$. 
To construct a proper 2-D Riemann flux formula, we first consider the two
limiting cases in which a propagating discontinuity may be described
by a 1-D Riemann flux formula.
\begin{enumerate}
\item For a discontinuity front perpendicular to the $x$ direction, where
$\w^{(,L)}=\w^{(,R)}$, one has
$$[\G_x(y)]_{\jp}={1\over 2}[\Omega(\w^{(R,)})+\Omega(\w^{(L,)})$$
\be
-\D_x^{(7)}(\ws)\cdot\delta_x\w_x](y)_{\jp},
\label{8a}
\ee
where $\D_x^{(7)}$ denotes the seventh row vector component of $\D_x$ matrix
and $\delta_x\w_x=(\w_x^{(R,)}-\w_x^{(L,)})$. The flux in
Eq.~(\ref{8a}) is extended to the range $y_k\le y\le y_{k+1}$
where the variables $\w^{(a,)}(y),\,a=L,R$
are continuous. We remind here that the upwind formula~(\ref{8a}) 
can be derived from the flux splitting introduced in Sect.~2.1, in the form
\be
\Omega^{(\pm,)}=\Omega(\w)\pm\D_x^{(7)}(\ws)\cdot\w_x,
\label{opm1}
\ee
which is equivalent to a local linearization of the $\Omega(\w(x))$ flux
around the $x=\xp$ point.
\item For a discontinuity front perpendicular to the $y$ direction, where
now $\w^{(L,)}=\w^{(R,)}$ the approximate Riemann solver reduces to
$$[\G_y(x)]_{\kp}={1\over 2}[\Omega(\w^{(,R)})+\Omega(\w^{(,L)})$$
\be
+\D_y^{(7)}(\ws)\cdot\delta_y\w_y](x)_{\kp},
\label{8b}
\ee
where $\D_y^{(7)}$ denotes the seventh row vector 
component of the $\D_y$ matrix and
$\delta_y\w_y=(\w_y^{(,R)}-\w_y^{(,L)})$. The $x$ range involved is
now $x_j\le x\le x_{j+1}$ where
the variables $\w^{(,b)}(x),\,b=L,R$ are continuous. Again,
the upwind formula~(\ref{8b}) may be derived from the splitting
\be
\Omega^{(,\pm)}=\Omega(\w)\mp\D_y^{(7)}(\ws)\cdot\w_y.
\label{opm2}
\ee
which represents a local linearization around the $y=\yp$ point.
\end{enumerate}

In the general case where a discontinuity front crosses 
the computational cell centered on $P$, an approximate Riemann solver can
be obtained by introducing a 2-D flux splitting. For that purpose,
we decompose each
$\Omega^{(\pm,)}$ components
in Eq.~(\ref{opm1}) along the $y$ direction, with the requirement
to have the same form of the symmetric decomposition
of the $\Omega^{(\pm,)}$ components in Eq.~(\ref{opm1}) along the $x$
direction. This compound flux splitting, when interpreted as an
approximate Riemann solver with local linearization
(thus implying to neglect $O(\delta_x\w_x\delta_y\w_y)$ terms),
results in the four-state flux formula 
$$\G(\w_P)=$$
$${1\over 4}[\Omega(\w^{(R,R)})+\Omega(\w^{(R,L)})+
\Omega(\w^{(L,R)})+\Omega(\w^{(L,L)})]_P$$
\be
-\frac{1}{2}[\D_x^{(7)}(\wss)\cdot\delta_x\ws_x-
 \D_y^{(7)}(\wss)\cdot\delta_y\ws_y]_P,
\label{8d}
\ee
where ($a,b=R,L$):
$$\ws_x^{(a,)}=\frac{1}{2}(\w_x^{(a,R)}+\w_x^{(a,L)}),$$
$$\ws_y^{(,b)}=\frac{1}{2}(\w_y^{(R,b)}+\w_y^{(L,b)}),$$
and $\wss=(\w^{(R,R)}+\w^{(R,L)}+\w^{(L,R)}+\w^{(L,L)})/4$.
The $\G(\w)$ numerical flux given in Eq.~(\ref{8d})
has now all the desired formal upwind properties both along the $x$ and
$y$ directions, and reduces correctly to the 1-D limiting cases
Eq.~(\ref{8a}) and Eq.~(\ref{8b}). 
The composition rule used here cannot be interpreted as a simple
arithmetic average of independent 1-D Riemann solvers. In fact,
for a discontinuity front with arbitrary slope angle around $P$,  
the 1-D flux formula, say Eq.~(\ref{8a}), still applies and 
gives the upwind contribution along the $x$ characteristic modes. Near
the $y=\yp$ point, $\G^{(,b)}_x\,(b=L,R)$ is now a two-state flux function
and upwinding has to be completed
by taking into account also the $-\A_y$ characteristic modes along
the orthogonal $y$ direction. By applying then 1-D flux upwinding as
in Eq.~(\ref{8b}) to the flux $\Omega(\w)=\G_x(\w)$ and 
by discarding quadratic
terms, one recovers Eq.~(\ref{8d}). This composition procedure taken 
in reverse order,
starting now from Eq.~(\ref{8b}), yields an identical result 
under the essential assumption of linearization.

Finally, it is worth noticing that the viscous (resistive) model equation 
for the numerical flux function $\G$, 
consistent with Eq.~(\ref{8d}), has the form
$$\Omega_v(\w)= \Omega(\w)-{1\over 2}[h_x\D_x(\w)\cdot\derx\w_x-
h_y\D_y(\w)\cdot\dery\w_y],$$
showing how the dissipative term generalizes the classical 
$\eta{\bf J}=\eta\nabla\times \bb$ term in Ohm's law of resistive plasmas.

Having now completed the construction of the $\G(\w)$
upwind flux for the induction equations, it is possible to
design the overall numerical procedure to integrate the
MHD system. We summarize here the main computational steps:

\begin{enumerate}
\item At each stage of the Runge-Kutta cycle, for given
$[\uu,A](t)$ data at time $t$, the averaged $[\hb_x,\hb_y]$
staggered fields are evaluated first by Eqs.~(\ref{bb}).

\item Using then the $(\uu,\hb_x,\hb_y)^T$ data,
all variables $(\uu,B_x,B_y)^T$ needed to compute the
fluxes defined in Eqs.~(\ref{f2}) are reconstructed at each
cell boundary point $(\xp,y_k)$ and conservative $x$-derivatives of
the $\ff$ flux can then be evaluated.

\item The complementary procedure, now to reconstruct
all variables $(\uu,B_x,B_y)^T$
for the fluxes defined in Eqs.~(\ref{g2}) at each $(x_j,\yp)$ point, 
gives the conservative $y$-derivatives of the $\gg$ flux.

\item A final reconstruction to the $(\xp,\yp)$ corner point
is needed to compute the $\G(\w)$ numerical flux in
Eq.~(\ref{8d}). This allows to integrate in time the vector magnetic
potential $A(t)$ by Eq.~(\ref{dAz}).
\end{enumerate}

The explicit representation of a divergence-free magnetic field
via a vector potential has, among others, the advantage of
an easy extension to
three-dimensional configurations. In fact, in this case one has
to discretize the constitutive relations
$$\bb=\nabla\times{\bf A},\quad \bb=[B_x,B_y,B_z]^T$$
in weak form to generalize Eqs.~(\ref{bb}). The $A_i,\,i=x,y,z$
components and the corresponding flux functions
$\Omega_i=[\v\times\bb]_i$ are now
located at the 3-D cell {\em edge} points
$P_i$, each centered along the corresponding $i^{th}$ 
coordinate but staggered with respect
to the remaining two directions. 
The evolution equation Eq.~(\ref{dAz})
readily generalizes to
$${d [A_i(t)]_{P_i}\over dt}=\Omega_i(\w_{P_i})$$
and the consistency arguments introduced in $2D$ case can
be applied to define the composition rules for
each $\Omega_i=\Omega_i(\w)$ scalar function  
since only two upwind directions, in turn, are now involved
({\em i.e.} each $\Omega_i$ involves a $2D$ Riemann solver). 

\subsection {Comments and discussion}

Some remarks are due in order to underline differences and
analogies with other proposed MHD schemes, in particular
with those presented by Dai \& Woodward (1998) (DW), 
by Ryu et al. (1998) (RY) and by Balsara \& Spicer (1999) (BS), 
all claiming to have ``divergence-free preserving
properties''. These works are based on second order
either Godunov or Roe-type schemes and a staggered discretization for
the induction equation of the form of our
Eqs.~(\ref{ind}) is used. 

At a second order level the averaged variables
$[\hb_x]_{\jp,k}$ and $[\hb_y]_{j,\kp}$ can be interpreted
as point valued variables at the same cell boundary points
(which we label hereafter as $(b_x,b_y)$ fields, to conform to
the DW and RY notation). In Sect.~3 we have shown that
these staggered, divergence-free variables, when used
in momentum and energy equations 
avoid the effects of numerical monopoles.
On the contrary, in all the cited works, while evolving in time the
staggered $(b_x,b_y)$ magnetic field, still
interpolation and upwinding procedures are based on 
cell centered $(B_x,B_y)_{j,k}$ variables and unwanted
compressive $\divb$ terms then necessarily set up.
Moreover, this ``duality'' in the magnetic field representation is
considered to be unavoidable in the Godunov-type Riemann
solvers formalism. For that reason, in the DW and BS papers
in particular, much attention has been
devoted to compare the different results produced by schemes advancing in time
only the $(b_x,b_y)$ staggered variables, where $(B_x,B_y)$
work as interpolated variables, and schemes where {\em also}
the $(B_x,B_y)$ cell centered components are (independently) evolved
(as for standard Godunov procedures for Euler equations).
Conclusions are mainly drawn at a qualitative level, and  
only in the DW paper some numerical results on the $\divb$ variable
constructed with the $(B_x,B_y)$ data are presented, showing
the onset of significant, even of $O(1)$ size, residuals.

In the present approach, we have demonstrated by analytical arguments that
the compressive components arise either in cases where cell centered fields
are evolved in time or they are simply given by interpolation.
In fact, at the leading second order interpolation used in the cited papers,
$$[B_x]_{j,k}={1\over 2}([b_x]_{\jp,k}+[b_x]_{\jm,k}),$$
$$[B_y]_{j,k}={1\over 2}([b_y]_{j,\kp}+[b_y]_{j,\km}),$$
one gets a $O(h^2)$ (for smooth fields) residual when the $\divb$ variable
is evaluated by centered first derivatives.
This compressive component associated to the $(B_x,B_y)$
fields cannot be considered to be small (of the
same order of the truncation error), since it is
easy to show that even for higher order ($r>2$) interpolation the 
leading $\divb=O(h^2)$ term never cancels out.

Our analysis shows that cell centered fields are not really needed in
upwind differentiation, even for Godunov-type schemes. In fact,
the MHD system structure relies on two different kinds of
magnetic field variables, depending on differentiation coordinates.
The first set, given by the $[b_x(x,),b_y(,y)]$ variables, which are 
continuous in the indicated coordinates, satisfy the divergence-free
condition and does not have a characteristic representation. 
The second one, given by $[B_x(,y),B_y(x,)]$, enters
the characteristic space and can then have (transverse)
discontinuities. The former quantities
are advanced in time as staggered field data, while the latter can 
be reconstructed by interpolation 
either at a cell boundary or at a cell corner point.

As a last point, we comment here on the way the flux for the
induction equations is derived. In the DW and BS schemes, the $\Omega$
fluxes (or electric fields components) are constructed by a simple arithmetic
averages in space and for each 1-D upwind flux. This approach, which seems
reasonable for second order accuracy, is not consistent since it does not
reduce to the original 1-D fluxes when the discontinuity front propagates
along one of the coordinate axes. This drawback led BS to introduce a
rather empirical {\em switch} to select the dominant direction of
front propagation. On the other hand, RY derives a formally correct flux,
of the same form of our Eq.~(\ref{8d}), by splitting the flux
$\Omega=v_xB_y-v_yB_x$ into two independent components ($v_xB_y$ and $v_yB_x$),
and 1-D independent upwindings along the $x$ and $y$ direction, respectively,
have then been applied. However, this 
computational trick has no
physical support, since the two characteristics spaces, spanned by
the $\A_x$ and $\A_y$, in the original
MHD equations are both based on the complete $\Omega$ flux.

\section{Implementetion of a third order LF-CENO scheme}

We consider the 2-D MHD system in cartesian $(x,y)$ coordinates,
in the conservation form given by Eqs.~(\ref{euler3}), to integrate
the density, momenta and energy variables, and by Eq.~(\ref{dAz})
to integrate the vector potential. The updated, line-averaged,
magnetic fields
$(\hb_x,\hb_y)$ are defined, at each time step, by the geometrical 
relations Eqs.~(\ref{bb}). 

We specify the general procedure outlined in the previous Sect.~3
by choosing high order reconstruction algorithms based on convex-ENO
(CENO) method, a local Lax-Friedrichs (LLF) flux splitting for upwinding and
a time integration step using
a third order TVD Runge-Kutta scheme.

All the indicated numerical ingredients are well documented
and tested in problems described by the Euler system of gas-dynamics
(Shu \& Osher 1988, for time integration,  Shu 1997, for general
ENO reconstruction
and Liu \& Osher 1998, for CENO method).
It is then sufficient to detail here the specific 
procedures allowing to extend CENO schemes to the MHD system and
having relevance to the divergence-free properties of the
magnetic fields.

\begin{enumerate}
\item  Among high order reconstruction algorithms, the recently proposed
CENO method has the main computational advantage to avoid the time
consuming characteristic decomposition of state variables, which is
usually adopted in upwind schemes.

To achieve this property, one consider first a TVD (monotone)
second order accurate interpolant for the
scalar variable $w(x)$ with data
$\{w_j\}$, where $w$ denotes any component of the state vector $\w$.
In the $\xm\le x \le\xp$ range, three-point 
linear polynomials have the form
$$L^{(k)}_j(x)=w_j+{1\over h_x}[\Delta w]_{j+k}(x-x_j),\quad k=0,1$$
where $[\Delta w]_i=w_{i+1}-w_i$. In classical TVD schemes a unique
interpolant
$\Ls_j(x)=w_j+{1\over h_x}[\de w]_j(x-x_j)$
with slope $[\de w]_j$ is chosen using min-mod ({\em mm}) limiters
to assure nonoscillatory properties 
$$[\de w]_j= mm([\Delta w]_{j-1},[\Delta w]_j).$$
The $mm(a,b)$ function is defined, as usual, by
$$mm(a,b)=sign(a)min[|a|,|b|],\quad ab > 0$$
and $mm(a,b)=0$ otherwise.

Using the selected $\Ls_j(w)$ polynomial, the interpolated values at the cell
boundaries are then given by $[w^{(L)}]_{\jp}=\Ls_j(\xp)$, 
$[w^{(R)}]_{\jp}=\Ls_{j+1}(\xp)$.

At any point of discontinuity as well at any smooth
extrema of $w(x)$ where the first differences change sign, 
the TVD polynomial reduces to the constant state
$\Ls_j(x)=w_j$. Clipping to first order accuracy at a function
jump is unavoidable
in any polynomial based reconstruction, while higher
($r\ge 2$) accuracy in smooth ranges can always be achieved
by enforced TVD or ENO procedures. This improvement, however,
usually requires a preliminary decomposition of the $\Delta\w$ differences
into characteristic modes, locally at each grid point. In the
CENO method a higher order interpolation is maintained only at
smooth regions while first order polynomials are used at the
function jumps, neither cases requiring 
a characteristic decomposition.

For third order reconstruction, in particular,
one has at disposal three
quadratic interpolants in the $\xm\le x\le\xp$ range
$$Q^{(k)}_j(x)=w_i+{1\over 2}([\Delta w]_i+[\Delta w]_{i-1})
{(x-x_i)\over h_x}$$
$$+{1\over 2}[\Delta^{(2)}w]_i{(x-x_i)^2\over h_x^2},$$
where $[\Delta^{(2)}]_i=[\Delta_i-\Delta_{i-1}]$
and centering refers to the index $i=j+k$ for $k=-1,0,1$. 
The CENO selection procedure allows
to construct the unique nonoscillatory interpolant
$$\Qs_j(x)=w_j+{1\over 2}[\de w]_j{(x-x_j)\over h_x}$$
$$+{1\over 2}[\dd w]_j{(x-x_j)^2\over h_x^2},$$
which is {\em closest} to the TVD lower order interpolant $\Ls_j(w)$.
This is obtained by computing the three differences
$$d^{(k)}_j=Q^{(k)}_j-\Ls_j,\quad k=-1,0,1$$
at the $x=\xm$ or $x=\xp$ boundary point. In a smooth range
all these distance indicators have the same sign and one can select then
$$\Qs_j=Q^{(k_0)}_j,\quad |d^{(k_0)}|=min_k|d^{(k)}|.$$
Only at a discontinuity point at least one indicator changes
sign and in this case one takes $\Qs_j=\Ls_j=w_j$ since
$\de=\dd=0$, clipping to a first order interpolation.

\item We apply then this reconstruction procedure, first
to each $u_{j,k}$ scalar component of the $\uu$ state vector,
to recover the point values
$$[u^{(L)}]_{\jp,k}=\Qs_j[u](\xp,y_k),$$
$$[u^{(R)}]_{\jp}=\Qs_{j+1}[u](\xp,y_k),$$
needed to compute the $\ff(\w_x,B_x)$ flux , and in a similar way
to recover the $u_{j,\kp}$ values needed to compute the 
$\gg(\w_y,B_y)$ flux.

\item
The definition of $(B_x,B_y)$ point values introduced in Sect.~3
requires the specification of the
nonoscillatory $\dd$ second derivative.
In the CENO framework one simply takes
$$\dd_x=mm[\Delta^{(2)}_{j-1},\Delta^{(2)}_{j},\Delta^{(2)}_{j+1}],$$
$$\dd_y=mm[\Delta^{(2)}_{k-1},\Delta^{(2)}_{k},\Delta^{(2)}_{k+1}].$$
We notice that this procedure
returns the smoothest among the indicated three second
numerical derivatives
if the stencil of the involved first differences 
$[\Delta_{j-2},....,\Delta_{j+1}]$ is monotone, while
$\dd=0$ if the first derivative has a jump or a smooth extremum.

For given $[\aB_x]_{\jp,k}$, the reconstruction of the related
divergence-free $[B_x]_{\jp,k}$ point values is given by the implicit
definition in Eqs.~(\ref{pval})
$$[B_x-\gm\dd_x B_x]_{\jp,k}=[\aB_x]_{\jp,k},\, \gm=1/24,$$
which we solve by using an explicit iteration procedure.
By setting $B^{(0)}_x=\aB_x$, at each $(\xp,y_k)$ point,
the sequence
\be
B^{(n)}_x=\aB_x+\gm\dd_x [B^{(n-1)}_x],\quad n=1,2,..
\label{seqx}
\ee
is clearly rapidly convergent since $\dd_x$ is at most an
$O(h_x)$ quantity. In a similar way, we compute $[B_y]_{j,\kp}$ by
iterating
\be
B^{(n)}_y=\aB_y+\gm\dd_y [B^{(n-1)}_y],\quad B^{(0)}_y=\aB_y.
\label{seqy}
\ee
In practical computations, for all the test problems presented in the
next section,
 we found that $n=5$ is sufficient to assure $\divb=0$ to within
machine accuracy both in the maximum and in the $L_1$ norm.

The computed $[B_x]_{\jp,k}$ point values computed in
Eq.~(\ref{seqx}) enter now the $\ff(\w_x,B_x)$
flux as they stand, while the $[B_y]_{\jp,k}$ component of the
$\w_x$ state vector needs a further interpolation. Using then
$[B_y]_{j,\kp}$ derived by Eq.~(\ref{seqy}), 
the cell centered $[B_y]_{j,k}$ field
is first reconstructed by taking
$$[B_y]_{j,k}={1\over 2}([\sB_y]_{j,\km}+[\sB_y]_{j,\kp}),$$
$$[\sB_y]_{j,\kp}=[B_y]_{j,\kp}-{1\over 8}[\dd_yB_y]_{j,\kp},$$
to be finally interpolated at the $(\xp,y_k)$ cell boundary point
like the other $u_{j,k}$ components of the $\w_x$ state vector.
It is worth noticing that the cell centered $[B_y]_{j,k}$ values 
have not divergence-free properties since the $\ff(\w_x,B_x)$ 
differentiation involves only the $x$ coordinate.

By symmetric arguments, the $[B_y]_{j,\kp}$ field enters the $\gg(\w_y,B_y)$
flux as it stands, while $[B_x]_{\jp,k}$ has now to be interpolated
using the cell centered values
$$[B_x]_{j,k}={1\over 2}([\sB_x]_{\jm,k}+[\sB_y]_{\jp,k}),$$
$$[\sB_x]_{\jp,k}=[B_x]_{\jp,k}-{1\over 8}[\dd_xB_x]_{\jp,k}.$$

\item The interpolated $[\w_x]_{\jp,k}$ or $[\w_y]_{j,\kp}$
are represented as two-point left-right values along the relevant $x$
or $y$ coordinate, and an approximate Riemann solver has then to be
specified to compute upwind fluxes. We have chosen the simple
 LLF flux composition, defined by
$$\ff(\w_x,B_x)={1\over 2}[\ff(\w_x^{L,},B_x)+\ff(\w_x^{R,},B_x)]$$
\be
-{1\over 2}\alpha_x(\ws_x)(\uu^{R,}-\uu^{L,}),
\label{lfx}
\ee
and
$$\gg(\w_y,B_y)={1\over 2}[\gg(\w_y^{,L},B_y)+\gg(\w_y^{,R},B_y)]$$
\be
-{1\over 2}\alpha_y(\ws_y)(\uu^{,R}-\uu^{,L}).
\label{lfy}
\ee
The scalar variable $\alpha_x(\ws_x)$ is given at each $(\xp,y_k)$ point
by the largest of the $\A_x$ matrix eigenvalues $\lambda_s(\ws_x)$,
and $\ws_x$ is the arithmetic average of the $\w_x$ left-right states.
In practice $\alpha_x=|v_x|+c_{f_x}$, where $c_{f_x}$ is the fast wave speed
along the $x$ direction. 
Correspondingly, $\alpha_y(\ws_y)=|v_y|+c_{f_y}$
gives the largest eigenvalue of the $\A_y$ matrix 
based on the arithmetic average of the left-right states
of $\w_y$ at the $(x_j,\yp)$ point.

\item Differences of flux values given in Eqs.~(\ref{lfx}),~(\ref{lfy})
provide  only second order accurate derivatives, even if the
reconstructed flux arguments share a higher accuracy order.
To keep third order in derivative approximations, we
construct the primitives
$$\aff_{\jp,k}=[\ff-\gm \dd_x\ff]_{\jp,k},$$
$$\agg_{j,\kp}=[\gg-\gm \dd_y\gg]_{j,\kp}$$
thus completing the integration scheme of Eqs.~(\ref{euler3})
for the six-component state vector $\uu_{j,k}$.

\item The $\Omega(\w)$ flux variable needs a proper upwinding
procedure, as shown in Eq.~(\ref{8d}). In a LLF scheme one has
$$\G(\w_P)=$$
$${1\over 4}[\Omega(\w^{(R,R)})+\Omega(\w^{(R,L)})+
\Omega(\w^{(L,R)})+\Omega(\w^{(L,L)})]_P$$
\be
-\frac{1}{2}[\alpha_x(\wss)\delta_x B_y-
 \alpha_y(\wss)\delta_y B_x]_P,
\label{lfa}
\ee
where all the arguments  $(v_x,v_y,B_x,B_y)$ are first  
interpolated at a common $P=(\xp,\yp)$ point.

\item Finally, for time integration, 
a three-step Runge-Kutta algorithm provides
the overall third order accuracy of the LF-CENO scheme.

\end{enumerate}

\section{Numerical results}

The proposed numerical problems are mainly concerned with the 
divergence-free property, which on numerical side entails two main aspects:
\begin{enumerate}
\item the existence of a
vector potential $A(\xp,\yp,t)$ as a continuous function at all
times, assuring regular field lines topology (only corners are allowed),
\item a vanishing $D_B\equiv[\divb]_{j,k}$ where field derivatives are computed
using the same field components and the same difference algorithms as in
dynamical flux calculations.
\end{enumerate}
Since this form of validation has no counterpart in other proposed
numerical works, comparisons with published data will cover necessarily
rather qualitative aspects. 
Beside the divergence-free condition, the numerical results give also
indications on the resolution properties, as well as
on the stability and reliability of the MHD code
described in the Sect.~4.

Finally, we remark that in our code the $A(x,y)$ vector potential
refers only to the nonuniform $(B_x,B_y)$ fields, since 
constant initial components are trivially preserved in time.
Therefore, for
problems having constant components $({B_0}_x,{B_0}_y)$, the
evolved $A(t)$ field is now defined by
$$B_x={B_0}_x+\dery A,\quad B_y={B_0}_y-\derx A,$$
replacing the original Eqs.~(\ref{b}) of Sect.~2.

\subsection{Shock-tube tests}

We first consider 1-D Riemann problems (using a full 2-D grid)
to check for resolution properties of the proposed
scheme. To that purpose, it is necessary to take into account
that high order schemes are not well suited
for shock-tube problems where lower order characteristic-based
schemes are optimal, instead.
  
We consider three problems documented
by Ryu and Jones (1995) in their Fig~1a, Fig~2a and Fig~5a, which we
here label RJ1, RJ2 and RJ3, correspondingly. In all the
indicated cases a uniform grid with $N_x=400$ grid points,
a grid size $L_x=1$,
an adiabatic index $\gamma=5/3$ and a CFL number $c=0.8$, are used.
In all numerical tests presented here, 
the parameters $\gamma$ and $c$ will always retain the same values.

In RJ1, the initial conditions for the state vector
$\w(x)=[\rho,v_x,v_y,v_z,B_y,B_z,p]^T$ are defined by
$$\w^{L}=[1,\,10,\,0,\,0,\,5B_0,\,0,\,20]^T,$$
$$\w^{R}=[1,\,-10,\,0,\,0,\,5B_0,\,0,\,1]^T,$$
and by a constant ${B_0}_x=5B_0$. Here left states refer to
$x <0.5$ and right states to $x>0.5$. The unit magnetic field is
$B_0=1/\sqrt{4\pi}$.
In Fig.~\ref{st1} the evolved variables $(\rho,p,B_y,v_x)$
are shown at time $t=0.08$, as in the referenced RJ paper.

In the RJ2 test initial conditions are defined by
$$\w^{L}=[1.08,\,1.2,\,0.01,\,0.5,\,3.6B_0,\,2B_0,\,0.95]^T,$$ 
$$\w^{R}=[1,\,0,\,0,\,0,\,4B_0,\,2B_0,\,1]^T,$$
and the constant magnetic field is now ${B_0}_x=2B_0$.
The evolved variables $(\rho,p,B_y,v_y,B_z,v_z)$
at time $t=0.2$ are shown in Fig.~\ref{st2}.

Finally, the RJ3 problem, with initial data
$$\w^{L}=[1,0,0,0,1,0,1]^T,$$
$$\w^{R}=[0.125,0,0,0,-1,0,0.1]^T,$$
and ${B_0}_x=0.75$, 
is illustrated in the Fig.~\ref{st3}, for $t=0.1$.
This is already considered a classical test, related to
the presence of a compound wave (Brio \& Wu 1988).

As can be seen, the plotted results reproduce well
all the main expected features and compare with the
corresponding results obtained with higher grid resolutions
and more elaborate Riemann solvers (Ryu \& Jones 1995,
Jiang \& Wu 1999). Postshock oscillations, which are always
produced in any shock-capturing scheme (Arora \& Roe 1996),
appear here with vanishing amplitudes behind the fast moving shocks
but have significant sizes near the slow shocks and near the
expansion wave on the right hand side of Fig.~\ref{st3}.
At present, to our knowledge, no general cure has been envisaged to suppress
entirely this unphysical wave noise, which can then only be reduced by
adding numerical viscosity. In this sense, 
the observed oscillations allow to
estimate the implicit numerical viscosity of our
CENO-LF scheme to be somehow intermediate between the lower order
TVD code of Ryu \& Jones (1995) and the Weno-LLF MHD code of
Jiang \& Wu (1999). Other limiters have also been tested (van Leer, {\em
superbee}, and so on), with no significant improvements.
In any case, we want to stress again the point that high order schemes not
based on characteristics decomposition, like our code, are not particularly
designed to handle Riemann problems, where a lower order scheme may be a
better choice.

To test the code for 2-D cases, we have run the previous Riemann
problems RJ1 and RJ2
with structures propagating along the main diagonal
of a computational box with sizes $L_x=\cos\alpha,\,L_y=\sin\alpha,$
where $\alpha=\pi/4$. In this way the diagonal has a unit size $L=1$
and $h=1/N_x$ is the size of the cell diagonals.
Initial conditions are then assigned to the state vector 
$\w(\xi)=[\rho,v_{\xi},v_{\eta},v_z,B_{\eta},B_z,p]^T$
along the coordinate $\xi=x\cos\alpha+y\sin\alpha$, with now
$B_{\xi}={B_0}_{\xi}$ being constant. 
Boundary conditions are specified by imposing the continuity of all variables
along the traverse direction $\eta=y\cos\alpha-x\sin\alpha$,
extended to the $x< 0,\, x> L_x$
and to the $y< 0,\, y> L_y$ sides. We used $N_x=N_y=256$ grid points 
and we found that this grid spacing is hardly sufficient to
recover the main flow structures.

The evolved variables $\w(\xi)$ are shown in
Fig.~\ref{st4} for the rotated 1-D Riemann problem of Fig.~\ref{st1},
and in Fig.~\ref{st5} for the rotated 1-D Riemann problem 
of Fig.~\ref{st2}, at corresponding times. 
The plotted results compare to the ones  
presented by Ryu et al.~(1998) for the same Riemann problems.
The fact that the small oscillations observed in the corresponding 1-D cases
are now less apparent, is due to the higher numerical dissipation produced
by the lower resolution.

In a 2-D shock-tube problem, the divergence-free condition can be
simply expressed by  a constant $B_{\xi}$ field, {\em i.e.} by
the $dB_{\xi}/d\xi=0$ relation along both the $\xi$ and $\eta$ coordinates.
However, if $B_{\xi}$ is 
constructed using the cell centered $[B_x,B_y]_{j,k}$
fields, this conservation law is poorly verified, as can be seen
in Fig.~\ref{st6}a, where the numerical derivative 
$\Delta_{\xi} B_{\xi}/h$ for the RJ2 test is plotted. 
On the other hand, using the vector potential, point values
of the $B_{\xi}$ field are properly defined by 
$$B_{\xi}={B_0}_\xi+{\partial A\over\partial\eta}$$
and the divergence-free condition results if $A$ (as well as the other
dynamical variables) do not depend on the $\eta$ coordinate. This is 
documented by the 2-D structure shown in
Fig.~\ref{st6}b. The corresponding numerical derivative
$${1\over h}\Delta_{\xi}B_{\xi}=
{\Delta_{\xi}\Delta_{\eta}A\over h^2}$$
has now a maximum size of $\simeq 10^{-5}$.

\subsection{Slow wave steepening and shock formation}

Nonlinear wave steepening from continuous initial data is
a main feature of compressible flows. In the MHD case this 
problem has also interesting astrophysical aspects for the
study of intermediate shocks (shocks coupled to expansion waves),
already encountered in the previous shock-tube test RJ3.
On the computational side, wave steepening is significant for
high order schemes, whose small numerical dissipation may
model weakly resistive plasmas. Reference results are given by 
 Wu (1987) for physical setting in resistive MHD and by
Dai \& Woodward (1998) and Jiang \& Wu (1999) for numerical
testing.

Here we consider initial data defined by a (smooth) slow
wave front propagating along the transverse $\xi$ axis, with a
slope angle $\alpha=\pi/6$ with respect to the $x$ axis.
The initial conditions are defined by the characteristic differential
equations (the prime here denoting a $\xi$ derivative)
$$\rp=-{B_{\eta}\Byp\over(a^2-c_s^2)},\, \prp=a^2\rp,\,
\qxp={c_s\over\rho}\rp ,\, \qyp={c_f\over a\sqrt{\rho}}\Byp,$$
relating $(\rho,\,p,\,q_{\xi},\,q_{\eta})$ to the
$B_{\eta}$ field along the $\xi$ coordinate. 
The variables $a,\,c_s,\,c_f\,$ denote
the sound, slow and fast wave speeds, respectively.
We choose as initial profile $B_{\eta}(\xi)=\sin(2\pi\xi)$ and
a cartesian box with $L_x=1/\cos\alpha$ and $L_y=1/\sin\alpha$,
so that periodic boundary conditions can be applied along the 
$x$ and $y$ cartesian coordinates.

In Fig.~\ref{sw1} the $\xi$ profiles of the
variables $(\rho,\,p,\,B_{\eta},\,v_{\xi}\,)$ are shown at time $t=1$, 
when a shock train is already formed.
The corresponding 2-D plots of
the pressure $p(x,y)$ and of the vector potential $A(x,y)$ are also
shown in the Fig.~\ref{sw2}, to check for
accurate $\eta$ independence of the flow variables. As for the previous
shock-tube test of Fig.~\ref{st6}, a vanishing numerical $\divb$ comes
from the $\dere A\simeq 0$ condition. In Fig.~\ref{sw3}a a surface
plot of the variable $D_B$ is also shown, giving a value
$|D_B|_{max}\simeq 10^{-4}$ for the residual numerical monopoles, while
the $B_{\xi}$ component based on $(B_x,B_y)_{j,k}$ values shows
a much higher derivative $O(10^{-1})$, as can be seen in the Fig.~\ref{sw3}b.

\subsection{The Orszag-Tang MHD vortex problem}

A well known model problem to study the transition to MHD turbulence
is provided by the so-called Orszag-Tang vortex, which has been extensively
studied in its compressible version (for low Mach numbers)
by many authors using spectral methods. For initial
Mach numbers $M\ge 1$
this is also a valuable test for upwind codes, and it has been used
for almost all the latest schemes (Zachary et al. 1994, 
Dai \& Woodward 1998, Ryu et al. 1998,
Jiang \& Wu 1999). The referenced Orszag-Tang 
system is defined by the initial conditions
$$v_x=B_x/B_0=-\sin 2\pi y,\quad v_y=\sin 2\pi x,$$
$$ B_y=B_0\sin 4\pi x,\quad p=(\beta/2)B_0^2 ,\quad \rho=\gamma p,$$
where  $B_0=1/\sqrt{4\pi}$ and $\beta=2\gamma$ for the usual
$\gamma=5/3$ value.
The initial flow is the given by a 
velocity vortex superimposed to a magnetic vortex,
with a common (singular) X-point, but with a different modal structure.
This configuration is strongly unstable, giving rise to a wide spectrum
of propagating MHD modes and shock waves (here the initial Mach number
is $M=1$), and to the transformation of the initial X-point to
a current-sheet triggering the reconnection process.

For this test we have chosen a unit grid $L_x=L_y=1$ with
$N_x=N_y=192$ collocation points. In Fig.~\ref{ot1} we present the 
pressure $p$ and potential $A$ isocontours at t=0.5, 
showing the good qualitative 
agreement with the other published works. In particular, 1-D profiles
of the $p(x)$ variable at $y=0.4277$ (upper plot) and at $y=0.3125$
(lower plot) are shown in Fig.~\ref{ot2} for a more detailed comparison
with the corresponding plots given, respectively, by 
Ryu et al. (1998) and Jiang \& Wu (1999).
The latter reference also contains some quantitative estimate of how
significant magnetic monopoles may affect the computed solutions, thus
producing numerical instabilities in a long time evolution.

The monopole distribution $D_B$ (not plotted here) show only 
a few enhanced values $\simeq 10^{-5}$, thus  assuring  vanishing
$\divb$ condition.  
in the long time computations we found no
evidence of negative pressure nor other unphysical behaviors.
To check this point in more detail, we plot the magnetic field lines
at $t=3$ in Fig.~\ref{ot3}, showing how the regularity of
fieldlines is well preserved in time. This essential feature
of the computed divergence-free magnetic field allows to reproduce
typical magnetic phenomena, like the topology change induced by a vanishing
resistivity (here modeled by a low numerical diffusivity).
In fact, by comparing the latter $A$ distribution with the former
shown in Fig.~\ref{ot1}, it is apparent how the initial magnetic islands 
around the $X$-point merge by reconnection.

\subsection{Strong blast wave in free space}

The following two numerical tests concern the formation and
propagation of strong MHD discontinuities in a 2-D domain. These
model problems are representative of many astrophysical
phenomena where the magnetic energy has relevant dynamical effects.
In numerical schemes having poor divergence-free properties, the
(possible) 
onset of spurious solutions and of a negative gas pressure is
clearly enhanced in these physical regimes, 
since the magnitude of numerical monopoles
increases along with the background magnetic pressure. This
problem has been discussed, in particular, by Balsara \& Spicer
(1999), where also some quantitative estimate of the numerical
$\divb$ produced by Godunov schemes has been documented.

The first test problem concerns the
explosion of a circular dense cloud in a magnetized, initially static region.
Here we take again a square domain with $N_x=N_y=192$ grid points.
Initial conditions are specified by filling a circular region
located at the center and radius $r_0=0.125$  with a hot gas 
having $p=100$.
The background static fluid is characterized by $\rho=p=1$ and ${B_0}_x=10$.

In Fig.~\ref{bl} we show the density $\rho$, the magnetic potential $A$, 
the magnetic pressure $(B_x^2+B_y^2)/2$ and the solenoidal variable $D_B$
distributions at time $t=0.02$, which is
already representative of the generated complex flow structure. In particular,
the plotted results show the well preserved initial axial symmetries
(around both the $y=0.5$ and the $x=0.5$ axis) as well as
the regularity of magnetic fieldlines.
As we can see, the resulting numerical $D_B$ variable
has an isolated peak with magnitude $10^{-3}$ and otherwise vanishing  
sizes $<10^{-5}$.

\subsection{The fast rotor problem}

In the Balsara \& Spicer (1999) work, a model problem to study
the onset and propagation of strong torsional Alfv\'en waves,
relevant for star formation, has been presented and analyzed.
Following this reference, we have runned the same problem
using a square unit computational box and  $N_x=N_y=240$ grid points.
For propagating structures not intersecting the box 
boundaries, periodic conditions can be applied.
Initial conditions are specified by a rapidly rotating cylinder (the rotor)
with center  at the $x=0.5,\, y=0.5$ point and radius $r=0.1$.
The rotor has (initial) density $\rho=10$, angular velocity $\omega=20$ 
and it is embedded in a static and uniform fluid with $\rho=p=1$ and
${B_0}_x=2.5/\sqrt{\pi}$.

The flow pattern evolved at time $t=0.18$ (just before the shocked flow
reaches the boundaries) is shown in Fig.~\ref{rot}, representing 
as in Fig.~\ref{bl} the space distribution of
the density $\rho$, the magnetic potential $A$, 
the magnetic pressure $(B_x^2+B_y^2)/2$ and the solenoidal variable $D_B$,
that keeps everywhere below $4\times 10^{-4}$. 
Even if a lower resolution than in
the referenced paper is here adopted, and no smoothing has been 
applied to the initial density and rotation velocity discontinuities,
the numerical results give
convincing evidence on how a higher order scheme provides accurate
and well resolved profiles even when strong discontinuities develop.

\section{Conclusions}

We have introduced a general method to adapt upwind schemes
developed for Euler system to the corresponding MHD system
in order to assure the divergence-free condition. The proposed
approach can be applied to existing MHD codes as well
as to any higher order extensions. 

The use of a staggered collocation for the magnetic field components
entering the $\divb$ variable and of the related magnetic potential $A$
are well known general premises to represent a numerical divergence-free
magnetic field at the second order accuracy level.
We have thus introduced proper algorithms to extend this representation
to higher orders and to formulate upwind flux derivatives
using only the divergence-free variables, in order to avoid the
onset of numerical monopoles in the momentum and energy equations.
Moreover, by taking into account consistency arguments, we have proposed
a new formulation for the upwind flux for the induction equations.

As an application, we have constructed 
a simple and efficient third order LF-CENO based MHD code running
in multidimensional systems.
This code appear well suited for many astrophysical
problems where, beside strong shocks, reconnection phenomena, 
complex wave patterns and turbulence develop, as confirmed 
by the several numerical tests here presented.

\acknowledgments
The authors would like to thank Marco Velli for many helpful discussions
and for his support in completing this work.

\clearpage

\figcaption[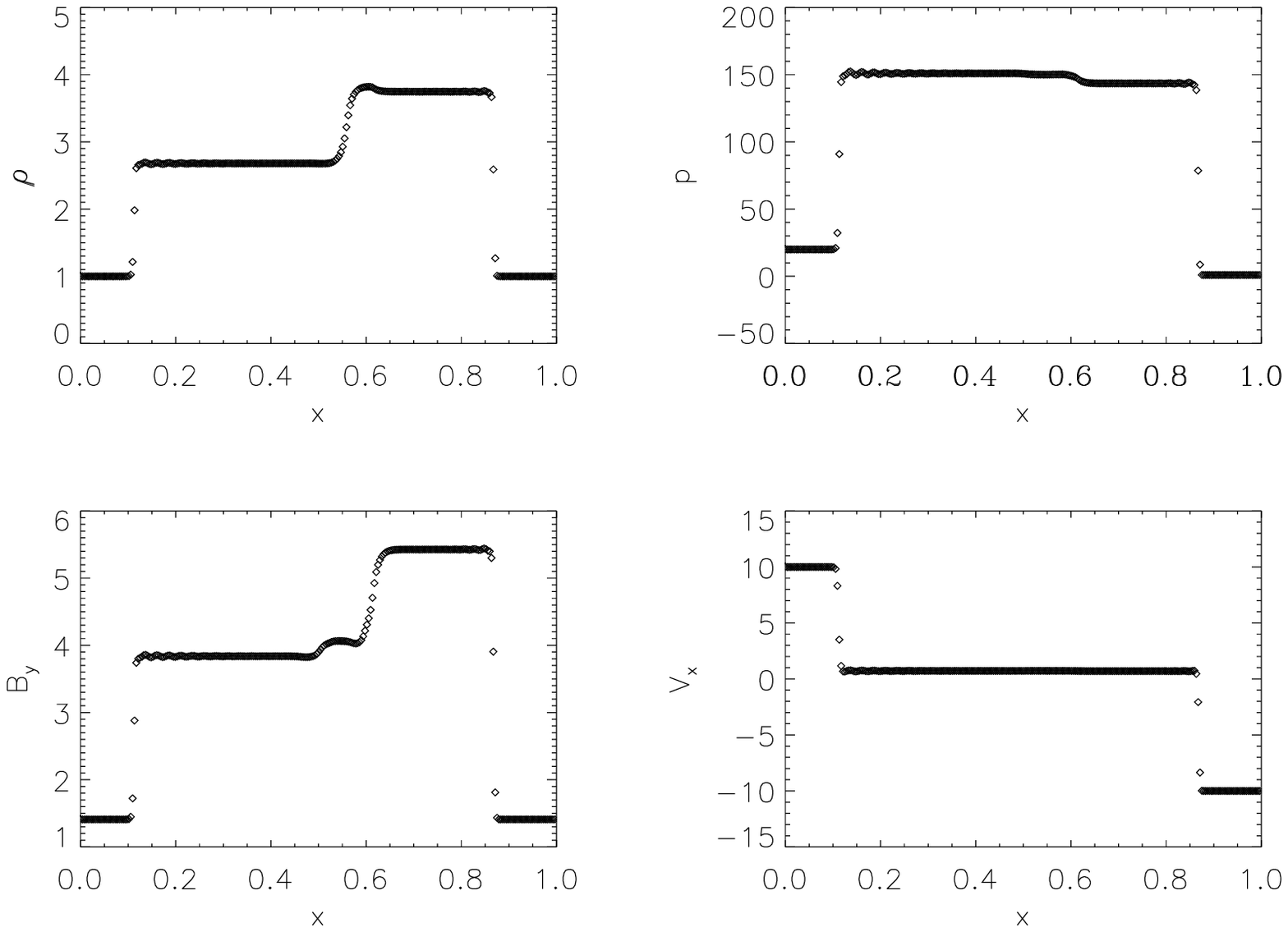]{
The indicated variables at time $t=0.08$
for the 1-D Riemann problem RJ1, using $N_x=400$ grid points.
\label{st1}
}
\figcaption[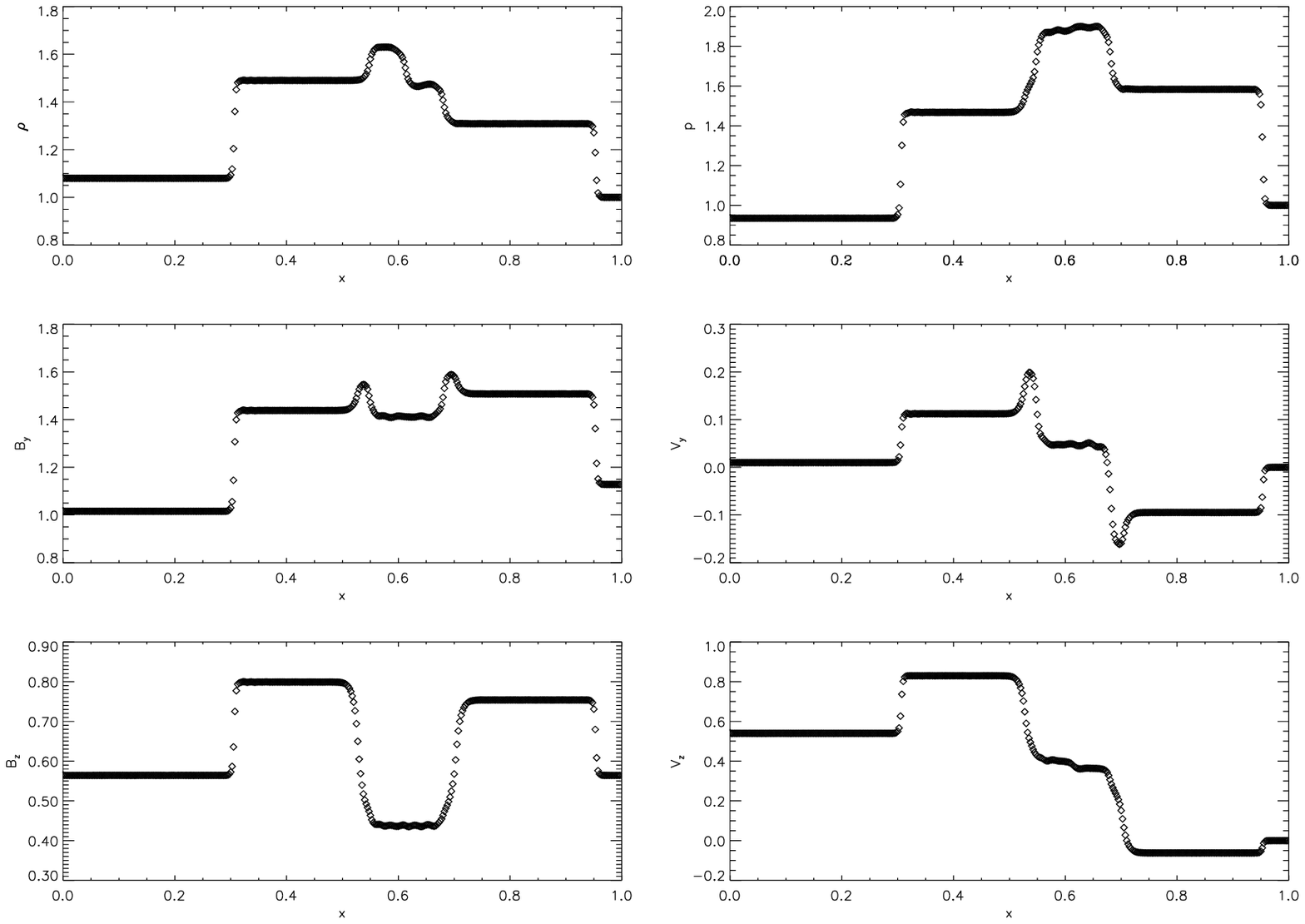]{
The indicated variables at time $t=0.2$
for the 1-D Riemann problem RJ2, using $N_x=400$ grid points.
\label{st2}}

\figcaption[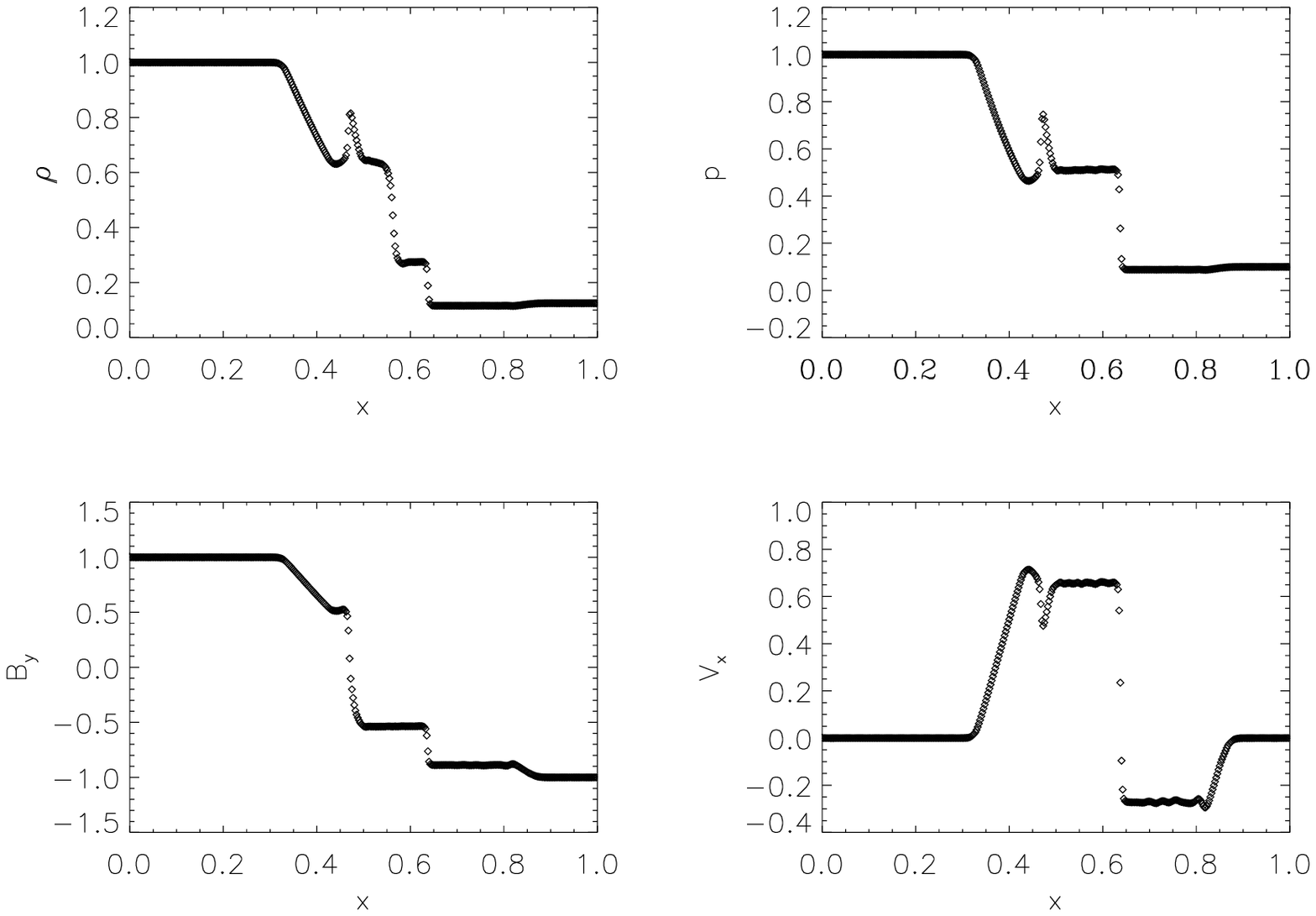]{
The indicated variables at time $t=0.1$
for the 1-D Riemann problem RJ3, using $N_x=400$ grid points.
\label{st3}}

\figcaption[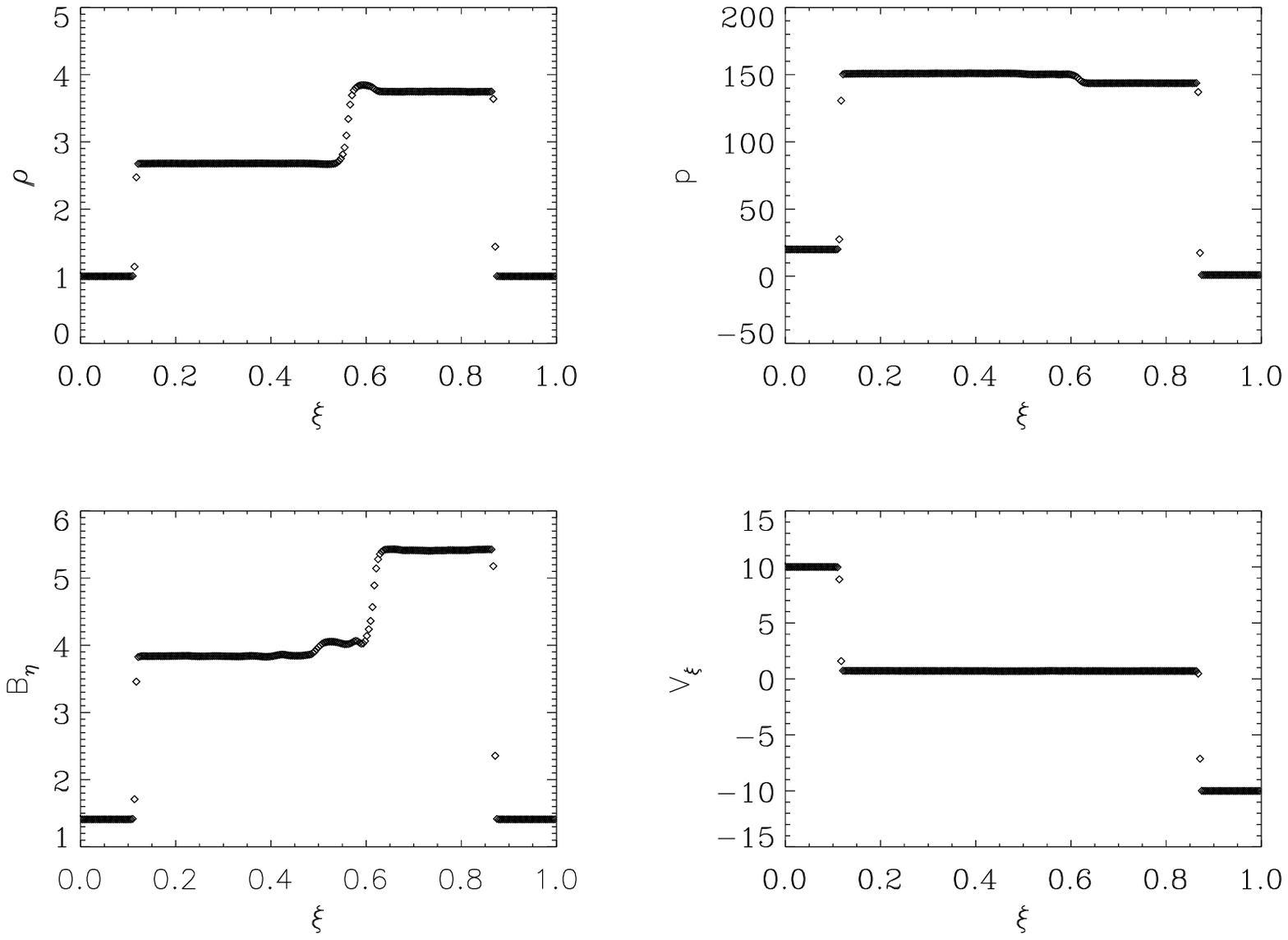]{
The same shock-tube problem as in Fig.~\ref{st1},
now along the main diagonal $\xi$ of a 2-D square computational box
with $(256\times 256)$ grid points.
\label{st4}}

\figcaption[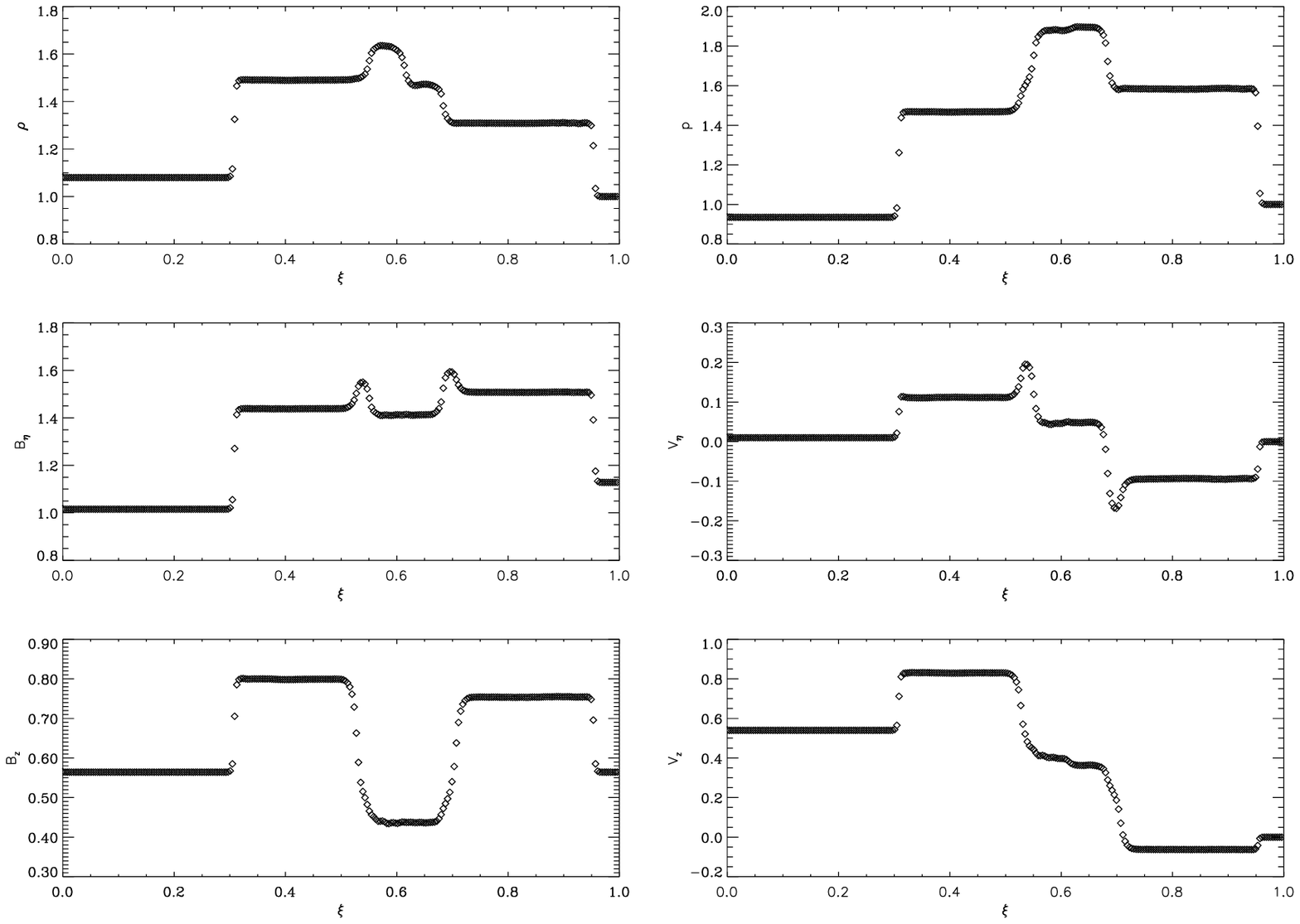]{
The same shock-tube problem as in Fig.~\ref{st2},
now along the main diagonal $\xi$ of a 2-D square computational box
with $(256\times 256)$ grid points.
\label{st5}}

\figcaption[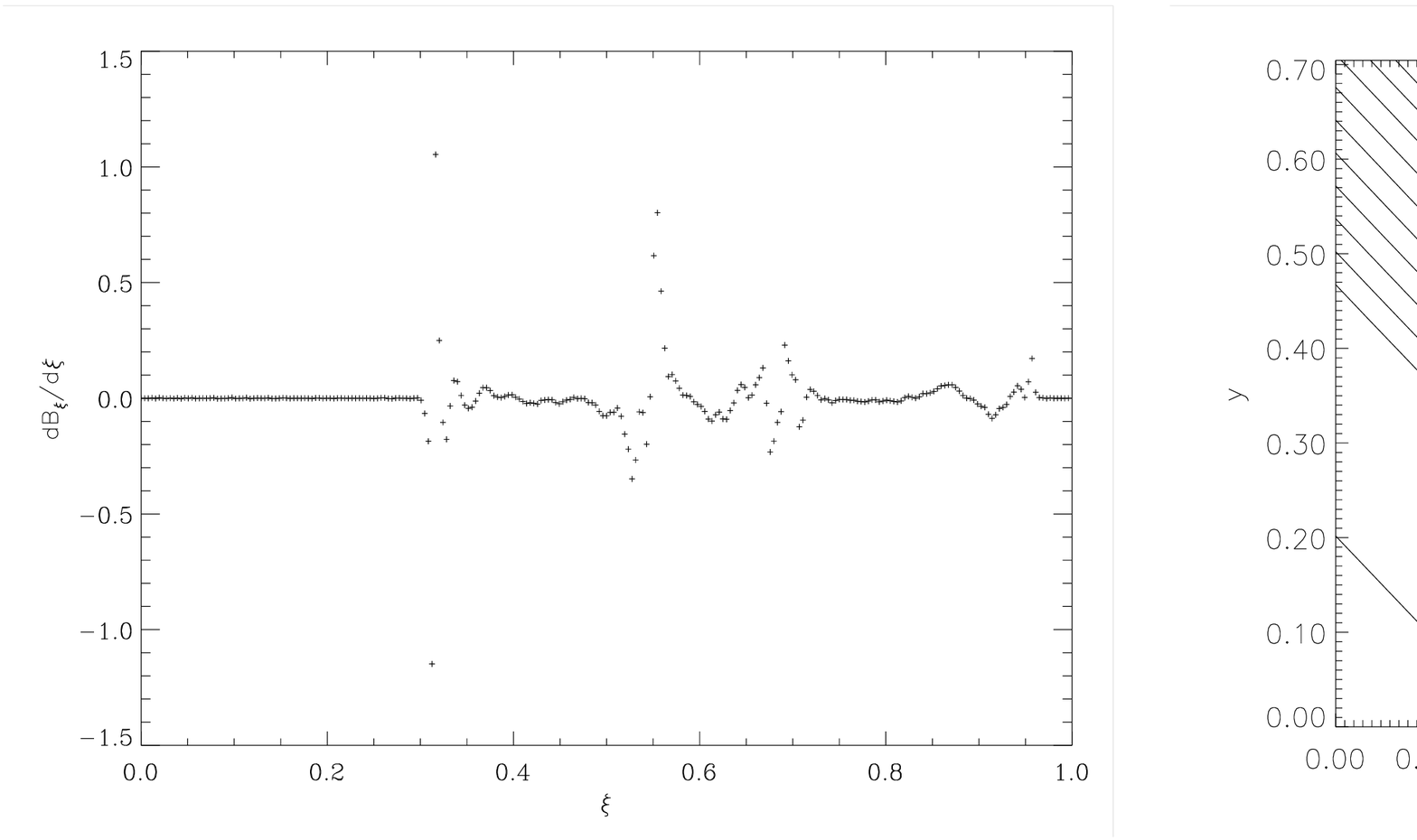]{
Left panel (a): the $\xi$ derivative of the $B_{\xi}$ field in
the 2-D shock-tube problem of Fig.~\ref{st5}. Right panel (b):
the corresponding isocontours of the magnetic potential $A$,
which clearly show the $\eta$-invariance.
\label{st6}}

\figcaption[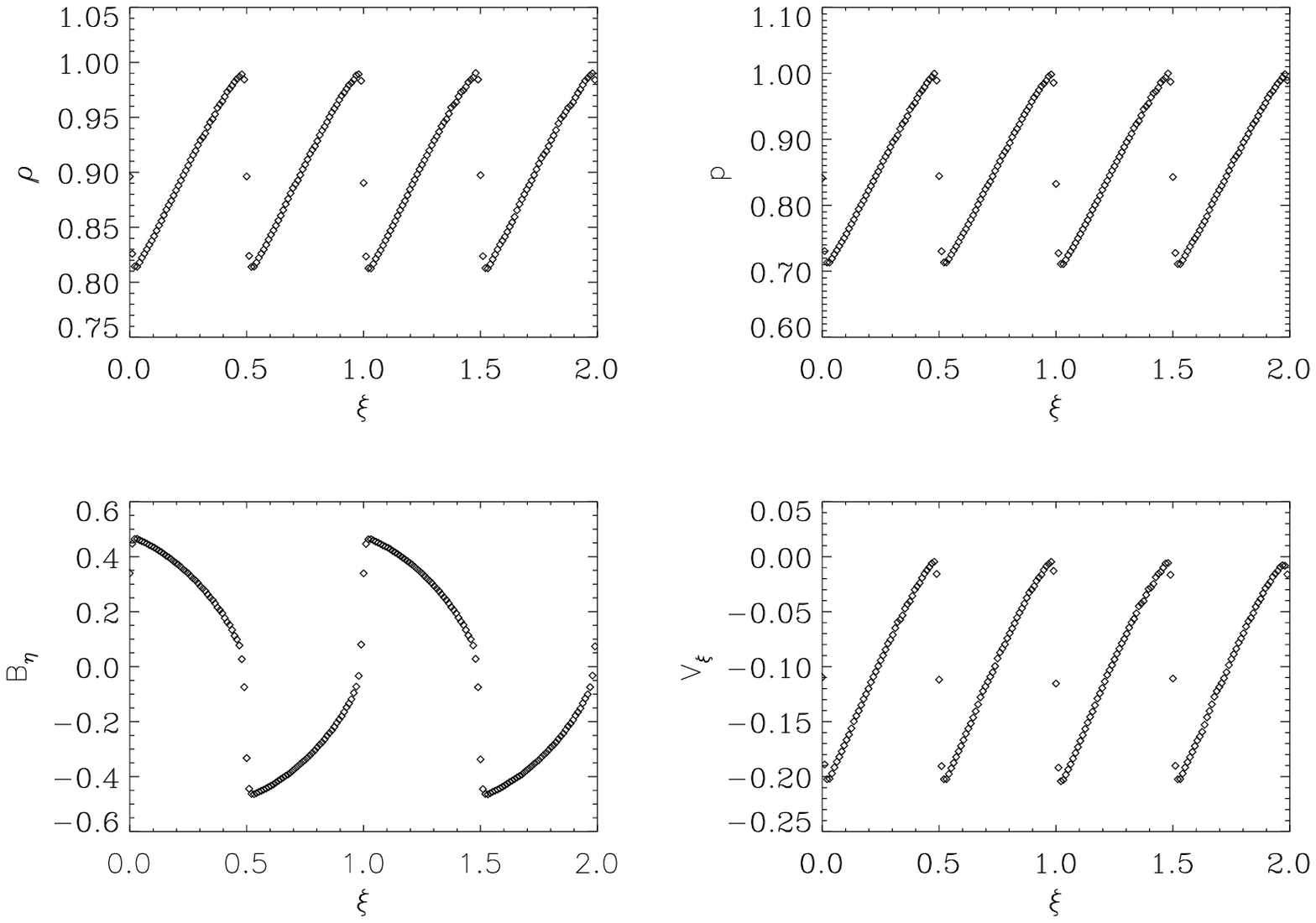]{
The indicated variables for the 2-D slow wave problem, 
along the $\xi$ coordinate and at time $t=1$. 
The slope angle is $\alpha=\pi/6$ and the computational box has
$(192\times 192)$ grid points.
\label{sw1}}

\figcaption[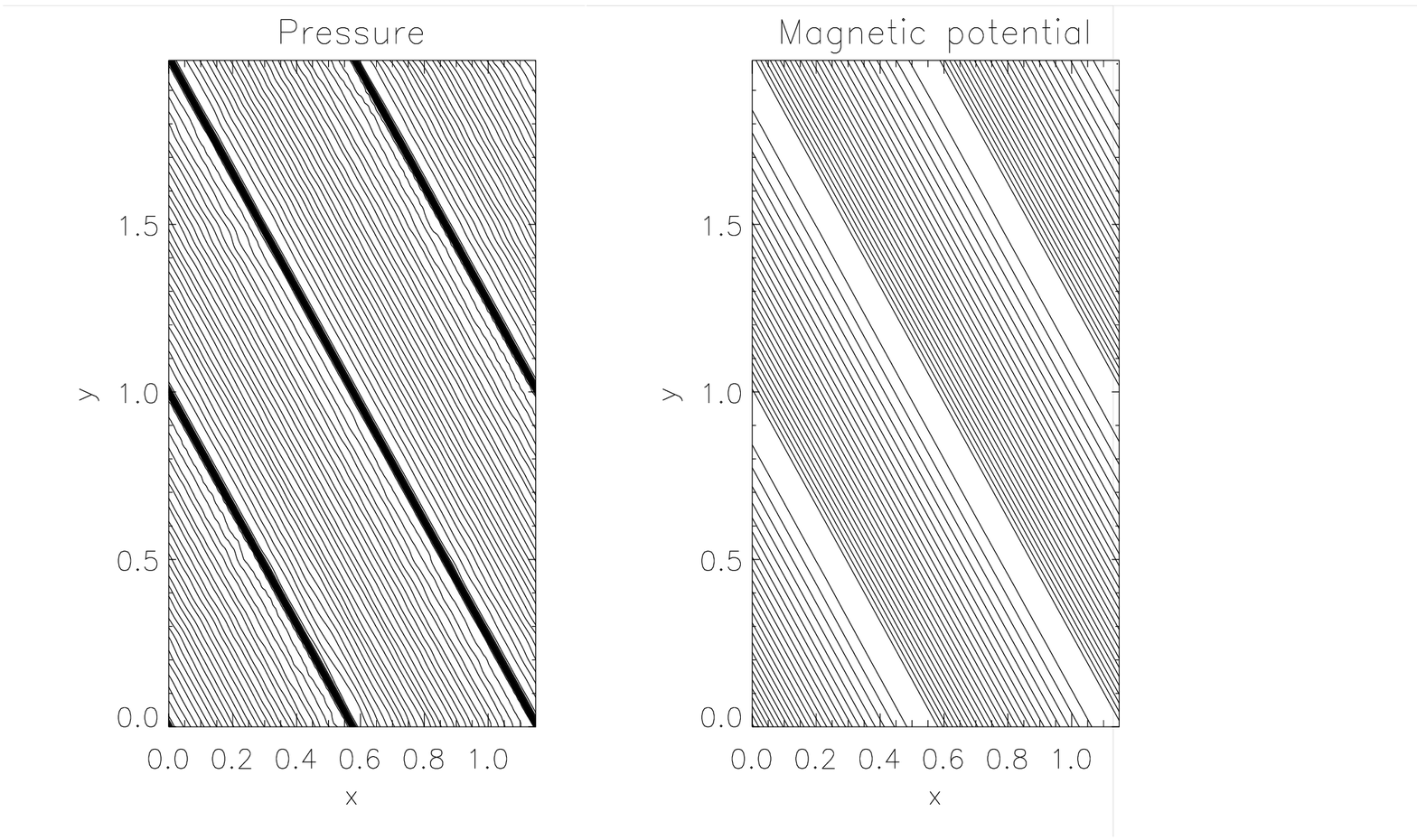]{
The pressure (left) and magnetic potential (right) isocontours 
for the slow wave problem of Fig.~\ref{sw1} at the same time $t=1$.
\label{sw2}}

\figcaption[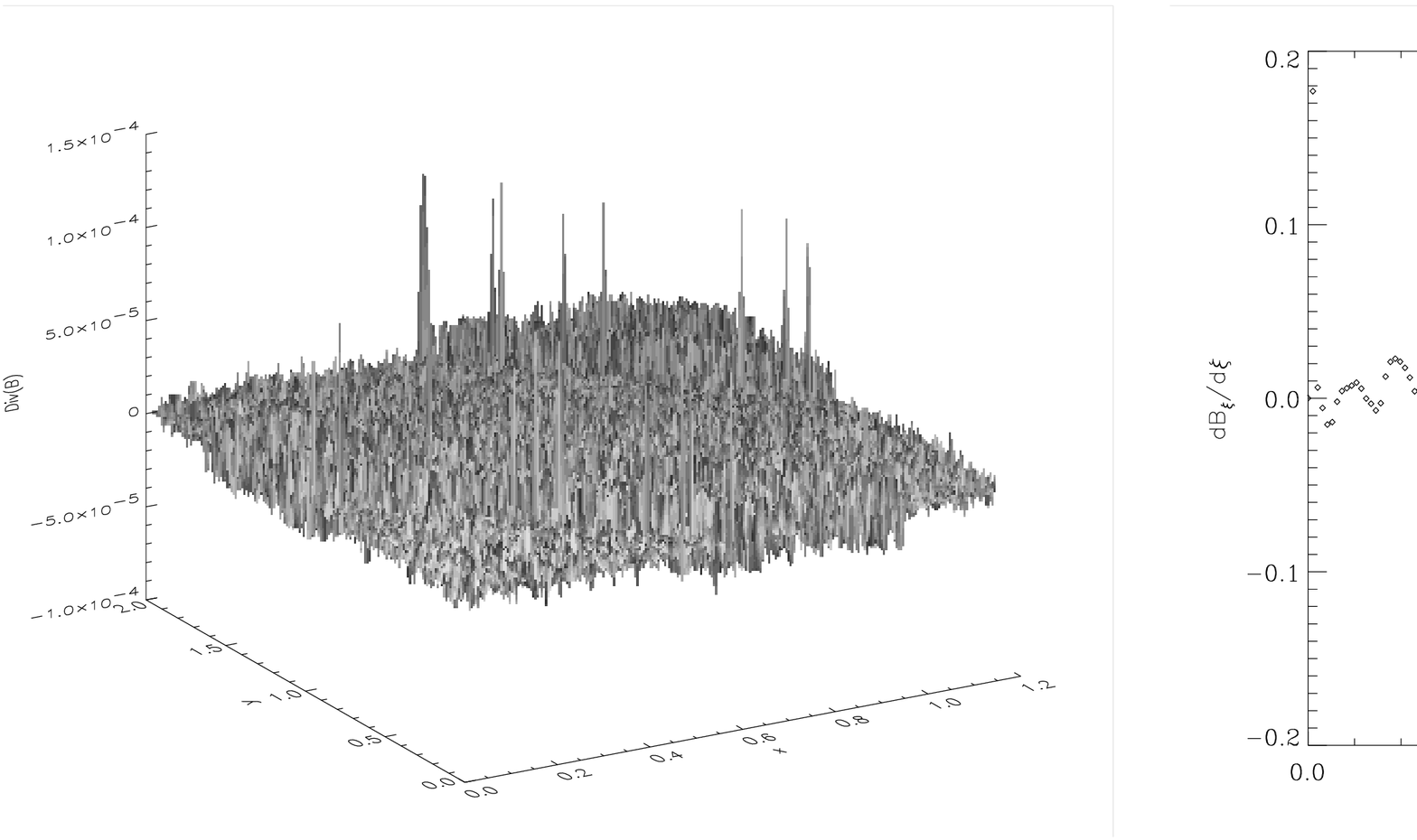]{
Left panel (a): A surface plot of $\divb$
for the slow wave problem at time $t=1$. Right panel (b):
The numerical derivative of the $B_{\xi}$ field, computed
with cell centered $(B_x,B_y)$ data.
\label{sw3}}

\figcaption[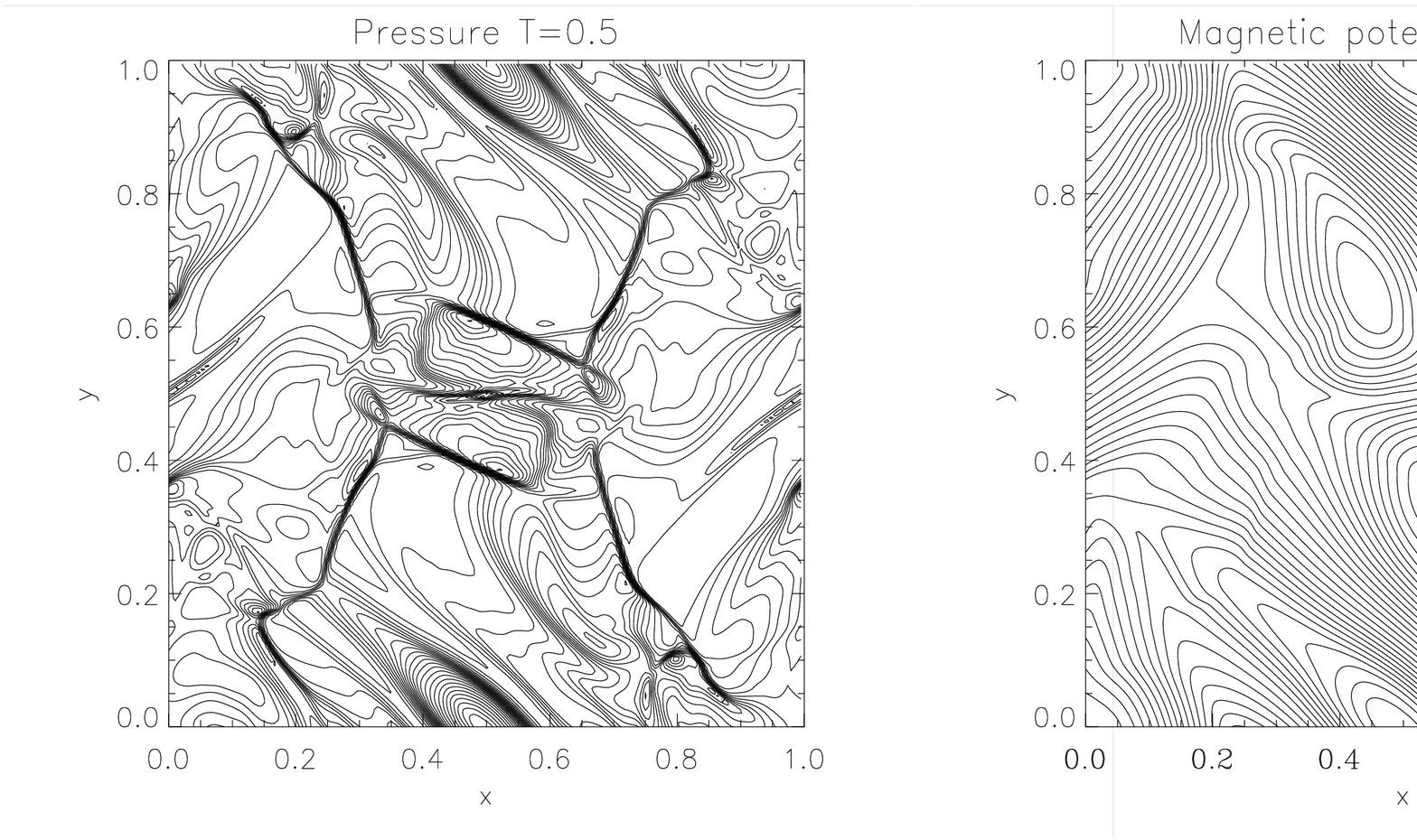]{
The pressure (left) and magnetic potential $A$ distribution (right)
in the Orszag-Tang problem at time $t=0.5$. The unit square computational box
has $(192\times 192)$ grid points.
\label{ot1}}

\figcaption[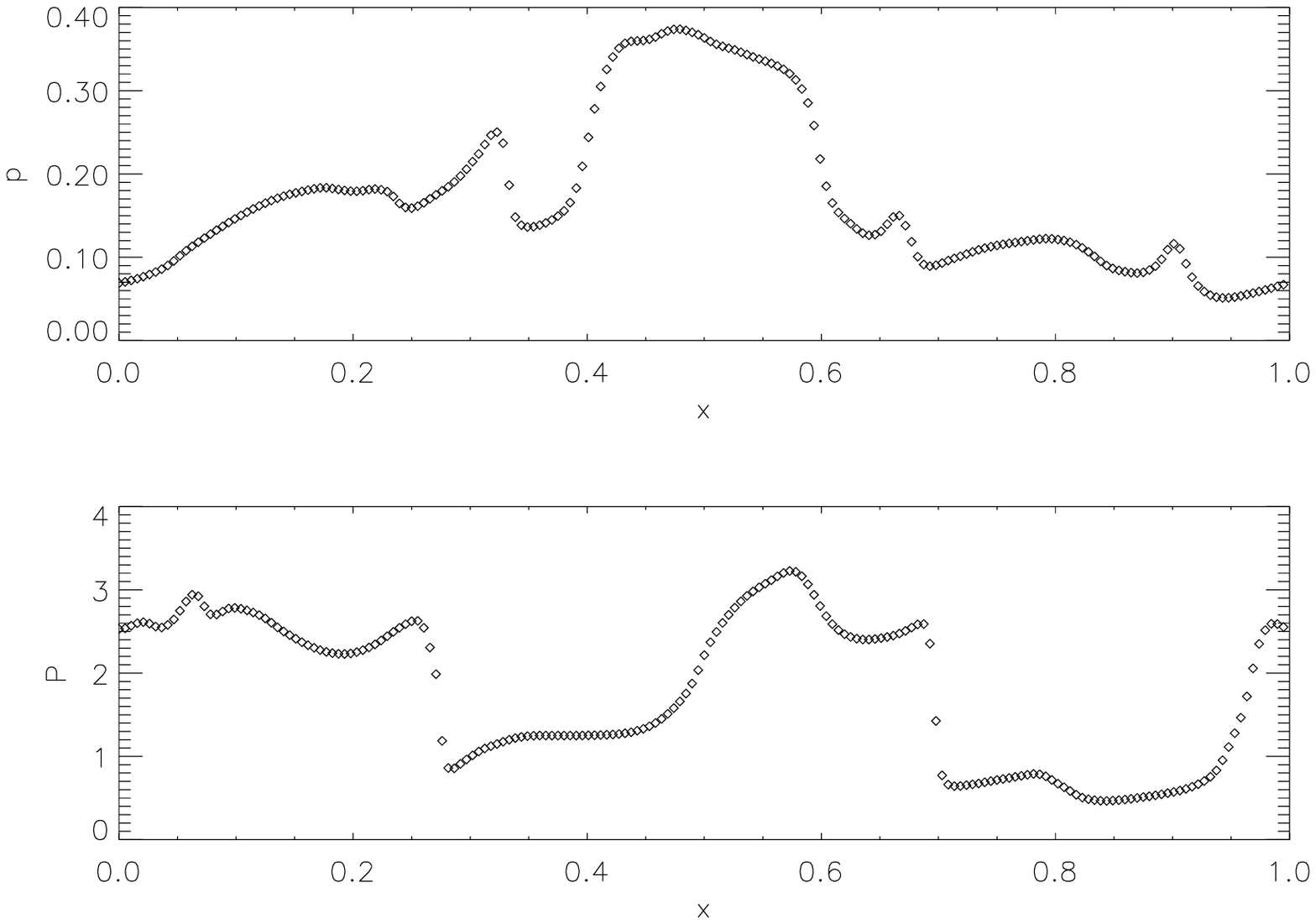]{
The 1-D pressure distribution
for the same problem as in Fig.~\ref{ot1} along a cut 
at $y=0.4277$ (upper plot) and at $y=0.3125$ (lower plot, where
a proper normalized pressure is shown, to compare
with the Jiang \& Wu (1999) data).
\label{ot2}}

\figcaption[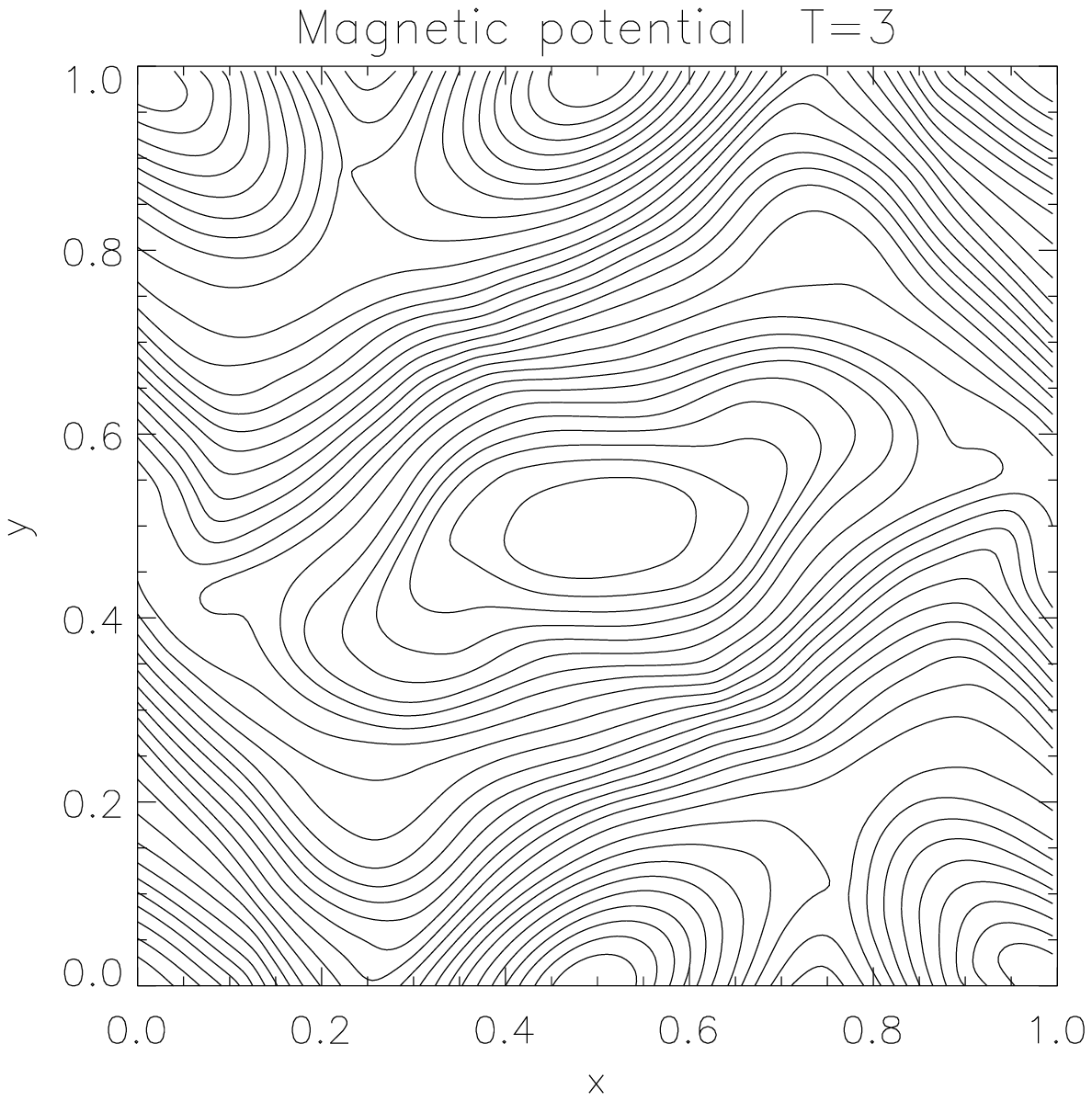]{
The magnetic potential $A$ distribution
for the same problem as in Fig.~\ref{ot1} at a later time
$t=3$. The reconnected central magnetic island is clearly shown.
\label{ot3}}

\figcaption[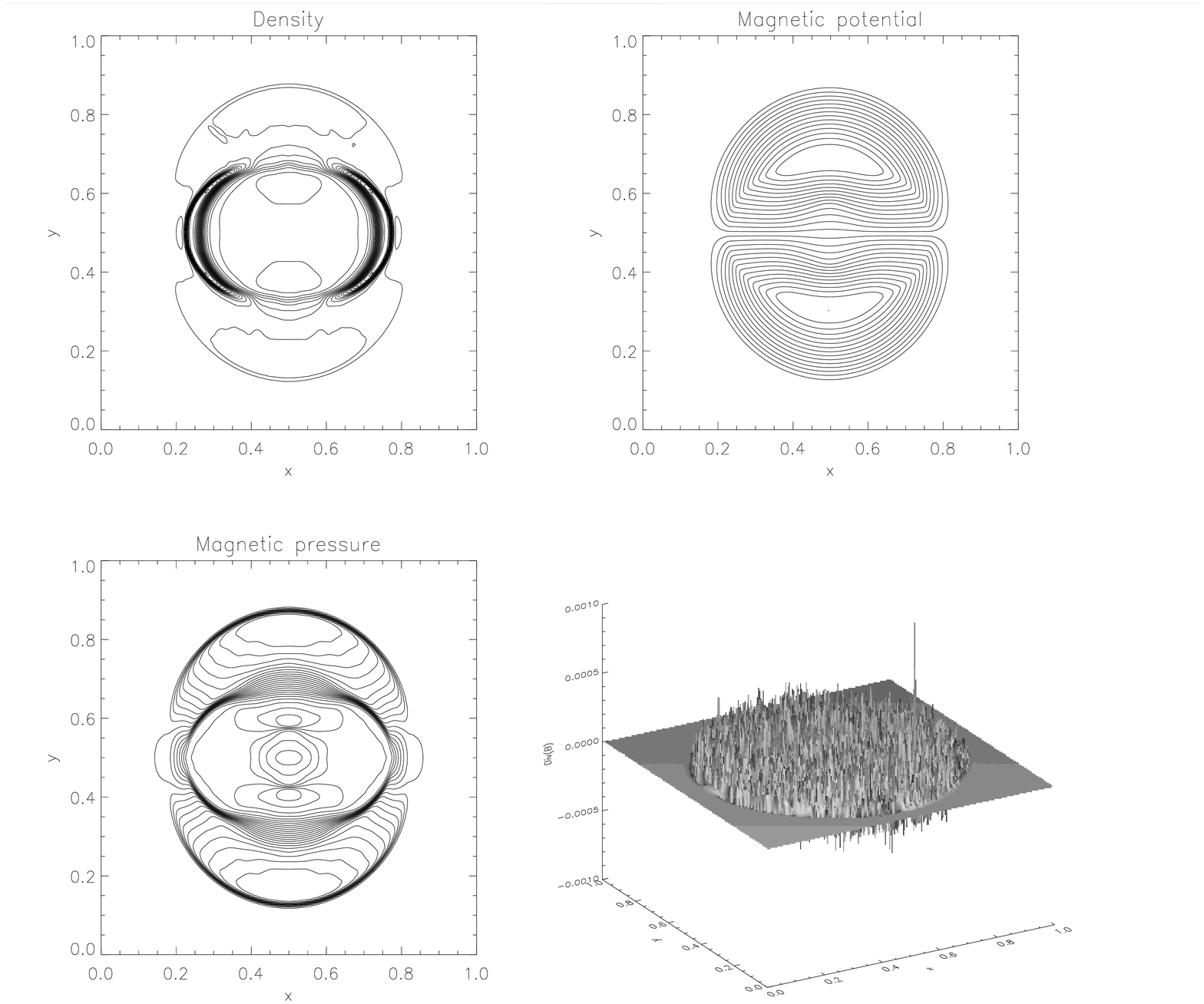]{
The indicated variables at time $t=0.02$ for the
blast wave problem. A unit computational box is used with
$(192\times 192)$ grid points.
\label{bl}}

\figcaption[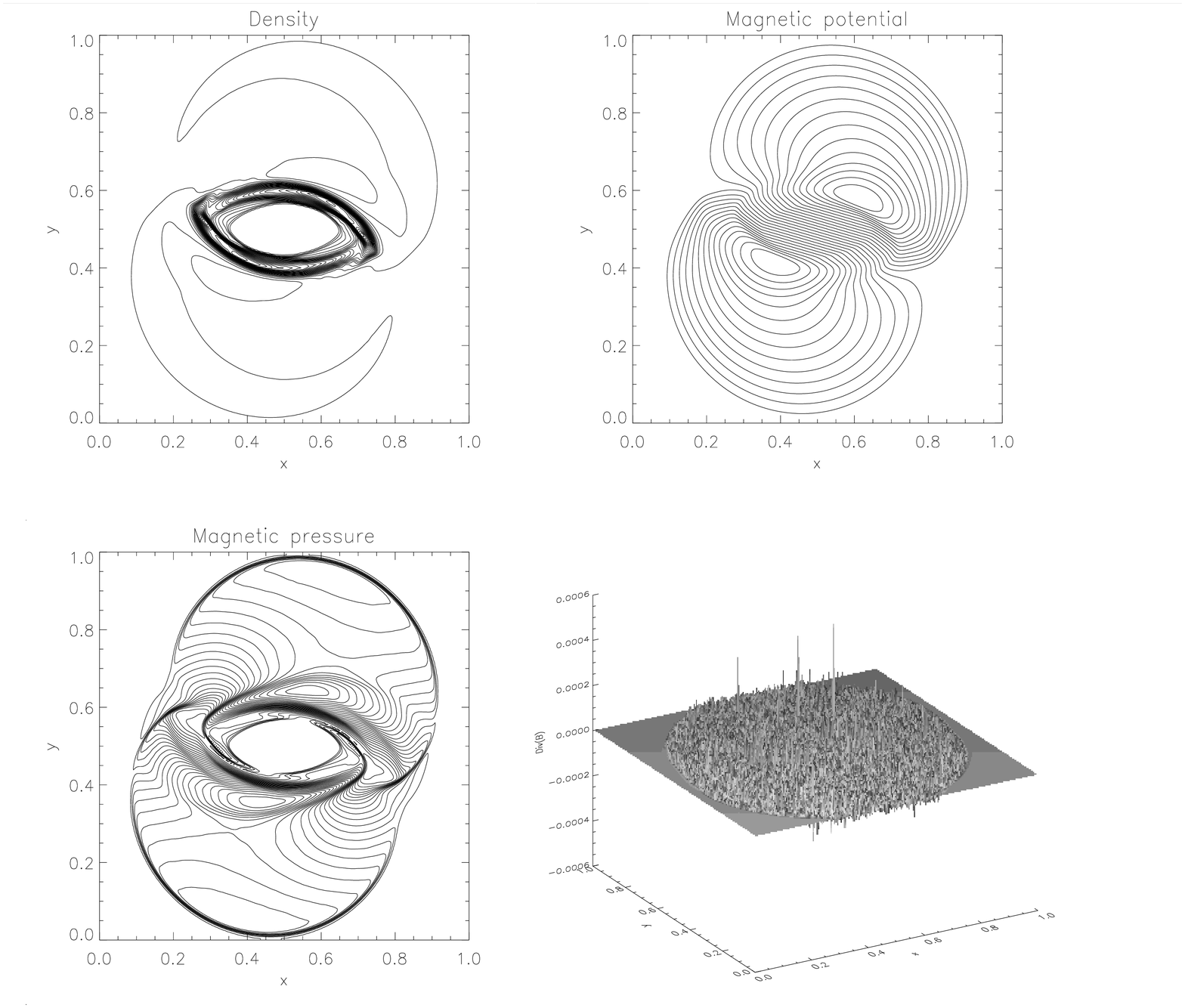]{
The indicated variables at time $t=0.18$ in the
rotor problem. A unit computational box is used with
$(240\times 240)$ grid points.
\label{rot}}

\onecolumn

\clearpage
\plotone{f1.ps}
\clearpage
\plotone{f2.ps}
\clearpage
\plotone{f3.ps}
\clearpage
\plotone{f4.ps}
\clearpage
\plotone{f5.ps}
\clearpage
\plotone{f6.ps}
\clearpage
\plotone{f7.ps}
\clearpage
\plotone{f8.ps}
\clearpage
\plotone{f9.ps}
\clearpage
\plotone{f10.ps}
\clearpage
\plotone{f11.ps}
\clearpage
\plotone{f12.ps}
\clearpage
\plotone{f13.ps}
\clearpage
\plotone{f14.ps}

\end{document}